\documentclass[onecolumn,twoside]{IEEEtran}

\usepackage{mathpazo}
\usepackage{times}
\usepackage{amsmath}
\usepackage{amsfonts}
\usepackage{latexsym}
\usepackage{amssymb}
\usepackage{cite}
\usepackage{upref}
\usepackage{theorem}
\usepackage{graphicx}
\usepackage{psfrag}
\usepackage{color}
\usepackage{setspace}

\newtheorem{proposition}{Proposition}

\newtheorem{theorem}{Theorem}
\newtheorem{lemma}{Lemma}
\newtheorem{definition}{Definition}
\newtheorem{corollary}[theorem]{Corollary}

\DeclareMathOperator*{\nn}{\nonumber}
\DeclareMathOperator{\E}{\mathbb{E}}
\DeclareMathOperator{\cC}{\mathcal C}
\DeclareMathOperator{\cX}{\mathcal X}
\DeclareMathOperator{\cY}{\mathcal Y}
\DeclareMathOperator{\cS}{\mathcal S}
\DeclareMathOperator{\cR}{\mathcal R}
\DeclareMathOperator{\cW}{\mathcal W}
\DeclareMathOperator*{\bp}{\bf p}
\DeclareMathOperator*{\bs}{\bf s}
\DeclareMathOperator*{\bpp}{{(\bf p)}}

\DeclareMathOperator*{\bpps}{{(\bf p, \bf s)}}
\DeclareMathOperator*{\CR}{\mathbf{CR}}
\DeclareMathOperator*{\regret}{\mathbf{Regret}}

\makeatletter
\def\blfootnote{\gdef\@thefnmark{}\@footnotetext}
\makeatother
 
\begin{document}

\title{Competitive Channel-Capacity}
\author{Michael Langberg, Oron Sabag}

\date{April 2022}

\maketitle
\begin{abstract}
      We consider communication over channels whose statistics are not known in full, but can be parameterized as a finite family of memoryless channels. A typical approach to address channel uncertainty is to design codes for the worst channel in the family, resulting in the well-known compound channel capacity. Although this approach is robust, it may suffer a significant loss of performance if the capacity-achieving distribution of the worst channel attains low rates over other channels. In this work, we cope with channel uncertainty through the lens of {\em competitive analysis}. The main idea is to optimize a relative metric that compares the performance of the designed code and a clairvoyant code that has access to the true channel. To allow communication rates that adapt to the channel at use, we consider rateless codes with a fixed number of message bits and random decoding times. We propose two competitive metrics: the competitive ratio between the expected rates of the two codes, and a regret defined as the difference between the expected rates. The competitive ratio, for instance, provides a percentage guarantee on the expected rate of the designed code when compared to the rate of the clairvoyant code that knows the channel at hand. Our main results are single-letter expressions for the optimal {\em competitive-ratio} and {\em regret}, expressed as a max-min or min-max optimization. Several examples illustrate the benefits of the competitive analysis approach to code design compared to the compound channel.
\end{abstract}

\blfootnote{M. Langberg is with the Department of Electrical Engineering at the University at Buffalo (State University of New York). 
Email: \texttt{mikel@buffalo.edu}.
O. Sabag is with the Rachel and Selim Benin School of Computer Science and Engineering, Hebrew University of Jerusalem, Israel. Email: \texttt{oron.sabag@mail.huji.ac.il}. The work of M. Langberg was supported in part by the US NSF under award CCF-1909451. A preliminary version of the current manuscript has been accepted for presentation at ISIT 2023.}

\section{Introduction}
The study of communication over channels whose statistics are not fully known can be modeled as a family of channels $\{W_s\}_{s \in \mathcal S}$ indexed by a set of possible states $\mathcal S$.
Dating back to the 1960's, several models for channel uncertainty have been studied: channels in which the state is fixed over time but unknown to the encoder and decoder, i.e., the compound channel, arbitrarily varying channels (AVC) in which the unknown state can change arbitrarily over time (perhaps subject to constraints), finite state channels (FSC) in which the channel state is governed by a known probability distribution, feedback channels that enable the encoder to learn the channel state, models in which the state is known at the encoder, and more, e.g., 
\cite{dobrushin1959optimum,blackwell1959capacity,wolfowitz1959simultaneous,root1968capacity,ziv1985universal,merhav1993universal,
csiszar2011information,
blackwell_capacities_1960,AhlswedeW:69correlated, ahlswede_elimination_1978,CsiszarN:88constraints,csiszar_capacity_1988,sarwate_robust_2008,bassily_causal_2014,
gallager1968information,wang1995finite,
shannon1958channels,costa1983writing,ahlswede1986arbitrarily,biglieri1998fading,verdu1994general,effros2010generalizing} and the survey in~\cite{lapidoth_reliable_1998}.

Traditionally, success criteria for communication in the presence of an uncertain channel model require the design of a single coding scheme that allows communication at a fixed rate no matter which channel state is realized; this approach forces code rates to be matched to the worst-case channel conditions, since higher rates cannot be accommodated in the worst-case state. Variations on this strategy require reliable decoding on all but a subset of channels, for example, with high probability over a random choice of the state, e.g., \cite{verdu1994general}. This fixed-rate approach is complemented by a variety of variable-rate  performance criteria. Variable-rate criteria no longer guarantee a fixed rate under (most or) all channel states; instead, they allow the rate to vary with the channel state in operation; definitions of reliability vary under different variable-rate models. {\em Rateless} codes, introduced in \cite{luby2002lt,mackay2005fountain,shokrollahi2006raptor}, achieve variable-rate coding by allowing the effective blocklength to vary with the channel state. They are useful for analysis and code design in both the finite-length and asymptotic regimes \cite{polyanskiy2011feedback}. 

In this work, we study communication in the face of uncertainty through the lens of {\em competitive analysis}.
In competitive analysis, one compares the achievable rates of solutions in which state information is not available with those in which state information is known in advance to all parties. The objective being the design of communications schemes that achieve rate which is {\em close} to that achievable when the channel state is known in advance and taken into account in code design.
Common metrics in this context include the {\em competitive-ratio} that measures the ratio between the rates achievable in the case of limited state knowledge and that with full state knowledge, and {\em regret} that measures the difference between the former and latter (expected) rates described above. The design of communication schemes with a competitive ratio $\alpha$ approaching 1 (or with regret approaching 0) guarantees that even in the face of uncertainty, the quality of communication matches the best possible under the given conditions. A competitive ratio $\alpha$ that is bounded away from 1 acts as a quality measure for the communication scheme at hand, guaranteeing that no matter what channel state is encountered during the communication process, the achievable rate will be guaranteed to be within an $\alpha$ multiplicative ratio of the best possible.

Competitive analysis, also referred to as ``online analysis'', has seen significant studies in the algorithmic literature in a variety of optimization problems in which constraints appear, and decisions are made, in a causal manner. Studied topics include for example job scheduling, self organizing data structures, network routing, caching/paging (or more generally the $k$-server problem), metrical task systems, resource allocation, covering and packing problems, and more, e.g., \cite{sleator1985self,sleator1985amortized,manasse1988competitive,borodin1992optimal,koutsoupias1995k,albers1999better,fiat2003better,pruhs2004online,borodin2005online,alon2009online,buchbinder2009design,bansal2015polylogarithmic}. 
In the area of communication and storage systems, competitive analysis has seen previous studies in the context of power management in energy harvesting or, alternatively, fading Gaussian systems \cite{merhav1998universal,sabag_filtering,buchbinder2010dynamic,ozel2011transmission,buchbinder2012dynamic,blasco2013learning,zhao2013competitive,vaze2013competitive,vaze2014dynamic,gomez2014competitive,zhao2015resource,mertikopoulos2015learning,stiakogiannakis2015adaptive,satpathi2016optimal,wei2016delay,mertikopoulos2016learning,yu2016dynamic,gomez2017competitive}, 
%in the context of point-to-point communication over unknown channel models, e.g.,  
%\cite{woyach2012comments}, 
%\cite{Shulman:09,shayevitz2005communicating, eswaran2007using, eswaran2009zero, shayevitz2009achieving, sarwate_rateless_2010,woyach2012comments,lomnitz2012communication, Blits:12,lomnitz2013universal,langberg2021beyond,joshi2022capacity}
and on other topics such as opportunistic network coding, opportunistic spectrum access in wireless systems, wear leveling in flash memory, estimation, and control \cite{ben2006competitive,chang2008competitive,hsu2012opportunistic,sabag2022optimal}.

\subsection{Our results and related work}
Our work studies communication over an unknown memoryless channel taken from a collection $\{W_s\}_{s \in \mathcal S}$ of channels using the framework of rateless codes. The channel in use is unknown in advance to the encoder and decoder. For a fixed message length $k$, our objective is to design a single rateless code $\mathcal C_k$ that optimizes the competitive metric under study, be it the competitive ratio ($\CR$) or the regret ($\regret$). Roughly speaking, a rateless code $\mathcal C_k$ has a competitive ratio of $\alpha \leq 1$ if, for any channel $W_s$ in use, successful decoding is achieved at time no more than $1/\alpha$ times the optimal decoding time had one used a code designed with knowledge of $W_s$. Stated in terms of rate, code $\mathcal C_k$ has a competitive ratio of $\alpha \leq 1$ if, for any channel $W_s$ in use, successful decoding is achieved at rate no less than $\alpha$ times the channel capacity. Here, in the variable-rate setting, the communication rate of a message of $k$ bits is defined to be $k$ divided by the expected decoding time. (All definitions are given in detail in Section~\ref{sec:model}.) Similarly, in the $\regret$ objective, we consider the minimization of the difference between the described above rates.

In this work, we present single-letter expressions characterizing the optimal achievable competitive metrics of $\CR$ and $\regret$. Our expressions are given in Theorems~\ref{th:main} and \ref{th:regret} of Section~\ref{sec:main} and include an optimization over $|\mathcal S|$ input distributions; one input distribution per channel state. Code design is obtained through a randomized construction involving the optimizing distributions above. Specifically, our optimal codes are not derived from a collection of random code symbols in which each entry is identically distributed. Rather, such constructions are shown to be in general suboptimal.
Accordingly, our code design uses a collection of $|\mathcal S|$ input distributions, where one switches between distributions at predetermined times that depend on the channel family at hand. Several examples are outlined in Section~\ref{sec:examples}.

\subsubsection{Related work}
While much of the theory for rateless codes relies on fixed, known channel statistics, most relevant to our work are 
a collection of results in the unknown and time-varying channel regimes, e.g., ~\cite{shulman2000static,Shulman:09,shayevitz2005communicating, tchamkerten2006variable,eswaran2007using, eswaran2009zero, shayevitz2009achieving, sarwate_rateless_2010,woyach2012comments,lomnitz2012communication, Blits:12,lomnitz2013universal,langberg2021beyond,joshi2022capacity}, in which rateless codes are used (with or without feedback) to obtain a communication rate that varies with the channel realization.\footnote{Several of the works above use a variant of rateless codes in which the blocklength is fixed while the message length is unlimited. In this setting, variability in rate is obtained by decoding a variable length prefix of the message. In Appendix~\ref{sec:app_comp} we compare the two rateless models showing that the rate region of the unlimited blocklnegth variant strictly includes that of the unlimited message-length variant; a comparison that may be of independent interest.}
A common theme among the works above is the design of codes whose rate for a given (unknown in advance) channel realization is tied to single-letter information theoretic quantities that vary with the realization at hand.
For example, in the context of arbitrarily varying channels \cite{blackwell_capacities_1960,ahlswede_elimination_1978}, or more generally the universal channel model \cite{lomnitz2013universal}, using feedback and common randomness \cite{eswaran2009zero,lomnitz2013universal} achieve a communication rate that may exceed the AVC capacity and equals the capacity of the time-averaged channel.

In this work, we study variable rate communication over an unknown channel $W_s$, chosen from a predefined collection of channels $\{W_s\}_{s \in \mathcal S}$, without the use of feedback or common randomness. Most relevant to the setting at hand are  \cite{shulman2000static,Shulman:09}  (or in a more general setting, \cite{langberg2021beyond}) that study the broadcast of a common message, termed {\em static broadcast}, using rateless codes to a collection of terminals $\{t_s\}_{s \in \mathcal S}$, each connected from the transmitter by the corresponding channel in $\{W_s\}_{s \in \mathcal S}$.
As different terminals can decode at different times, a rateless encoder and decoder imply variable rate decoding even given a common message.
Thus, in this setting, the static-broadcast rate-region $\cR$ includes achievable rate tuples $\{R_s\}_{s \in \mathcal S}$, i.e., rate $R_s$ for terminal $t_s$ connected from the transmitter by $W_s$.
In this context, an achievable tuple $\{R_s\}_{s \in \mathcal S}$ in the broadcast setting  corresponds naturally to  the achievable rate $R_s$ in point-to-point  communication over a ``partially" unknown channel $W_s$ from $\{W_s\}_{s \in \mathcal S}$, i.e.,  $W_s$ is unknown to the encoder but known to the decoder.
% Denoting the capacity of $W_s$ by $C_s$, roughly speaking, the competitive ratio for the channel family $\{W_s\}_{s \in \mathcal S}$ can now be implicitly derived from $\cR$ by
% $\CR = \max_{\{R_s\}_{s \in \mathcal S} \in \cR}\min_{s \in \mathcal S} {R_s}/{C_s}.$
% Similarly, $\regret = \min_{\{R_s\}_{s \in \mathcal S} \in \cR}\max_{s \in \mathcal S} {C_s}-{R_s}.$
In \cite{shulman2000static,Shulman:09}, the rate-region $\cR$ is described using a variant of an $n$-letter (for unbounded block length $n$) information-theoretic expression (the example of  $|\mathcal S|=2$ is further discussed in \cite{shulman2000static,Shulman:09} to obtain a single letter characterization of $\cR$ in that case). 
The rate-region $\cR$ can be used to derive competitive metrics; here, for the setting of state-information known only to the decoder. 
For example, roughly speaking, denoting the capacity of $W_s$ by $C_s$, 
the corresponding competitive ratio is $\max_{\{R_s\}_{s \in \mathcal S} \in \cR}\min_{s \in \mathcal S} {R_s}/{C_s}$.
% measures the optimal ratio between the .roughly speaking, the competitive ratio for the channel family $\{W_s\}_{s \in \mathcal S}$ can now be implicitly derived from $\cR$ by
% $\CR = \max_{\{R_s\}_{s \in \mathcal S} \in \cR}\min_{s \in \mathcal S} {R_s}/{C_s}.$
% Similarly, $\regret = \min_{\{R_s\}_{s \in \mathcal S} \in \cR}\max_{s \in \mathcal S} {C_s}-{R_s}.$

Beyond the fact that, in our study, channel state is unknown to both the encoder and decoder,\footnote{Later we show in our achievability that decoder state knowledge does not impact the achievable rates.} the work at hand differs from that of \cite{shulman2000static,Shulman:09}  (and the additional works mentioned above) in its explicit focus on a single-letter characterization of the competitive metrics of the channel family $\{W_s\}_{s \in \mathcal S}$. As discussed above, we present  $\CR$ and  $\regret$ as an optimization over $|\mathcal S|$ distributions, one corresponding to each channel in $\{W_s\}_{s \in \mathcal S}$. Our definitions, given in Section~\ref{sec:model}, follow those common in the finite-blocklength literature, e.g., \cite{polyanskiy2011feedback}, and thus our definition of rate slightly differs from that in \cite{shulman2000static,Shulman:09} in the sense that we determine the achievable rate using an expected decoding time. Our results hold for the rate model of \cite{shulman2000static,Shulman:09} as well. It is interesting to note that \cite{shulman2000static} also studies the family $\{W_s\}_{s \in \mathcal S}$ of {\em all} binary input channels, and shows that codes designed using the uniform distribution achieve a (surprisingly high, albeit potentially suboptimal) competitive ratio of $0.94$. However, this phenomenon does not hold for general inputs, as we later present, in Section~\ref{sec:examples}, example channels (with $|\cX|=4$) for which a code randomly designed by any single distribution (including the uniform one) has competitive ratio at most $0.5$ whereas the optimal competitive ratio is approximately $1$.

\section{Model: Competitive analysis for communication}
\label{sec:model}
 We consider a family of memoryless channels $W_s(y|x)$ that are parameterized with state $s\in[1:|\mathcal S|]$. The channel to be used among the $|\mathcal S|$ channels is unknown to the encoder and the decoder so that the code design is oblivious to the true channel in the given family. At the the high-level, our objective is to design codes that \emph{perform well} for all possible channels in a competitive manner. 

The alphabets of the channel input, output and the state are assumed to be finite, i.e., $|\cX|,|\cY|,|\mathcal S|<\infty$. The message $M$ is uniformly distributed over $\mathcal M \triangleq [1:2^k]$. Note that message size is not parameterized with a blocklength or rate. Instead, we consider rateless communication where the fixed-size message $M$ should be decoded with the least number of channel uses. That is, the decoder observes a stream of channel outputs and decides at each time if it wants to proceed with the communication or to abort communication and decode the message $M$.

\subsection{Code}
For a fixed number of bits, $k$, a rateless code $\mathcal C_k\triangleq(E,\{D_i\},\{H_i\})$ is defined by three mappings. The first is a sequence of decoder decisions functions
\begin{align}
    H_i: \mathcal Y^i \to \{0,1\}, \ \ i=1,2,\dots
\end{align}
If the decoder decides to decode at time $i$, $H_i(Y^i)=1$, the message is decoded with the mapping 
\begin{align}
    D_i: \mathcal Y^i \to  \mathcal M.
\end{align}
In all other times (i.e., $H_i=0$), we can arbitrarily define $D_i(Y^i) = 1$.
The encoder is defined by the mapping 
\begin{align}
 E: \mathcal M \to \mathcal X^\infty.  
\end{align}
The stopping time is defined as
    \begin{align}
        \tau_k = \min_i \{i: H_i(Y^i) =1 \}.
    \end{align}
For a fixed code, probability of error for a channel $s\in\mathcal S$ is   
\begin{align}
\label{eq:error}
    P_e(s)&\triangleq \Pr(M \neq  D_{\tau_k}(Y^{\tau_k})|S=s), 
\end{align}
and $P_e = \max_s P_e(s)$ denotes the maximal error among all channels.

The expected decoding time for channel  $s\in\mathcal S$ is denoted by 
\begin{align}
    \tau_s(k)&= \E[\tau_k|S=s].
\end{align}

To define the competitive metrics, as our superior yet impractical baseline, we choose codes where the encoder and the decoder mappings have access to the channel in use. We denote by $\mathcal C^\ast_k$ the set of these clairvoyant codes, and their corresponding expected stopping time as $\tau_s^\ast(k) = \mathbb{E}_s[\tau_k^\ast]$.

We can now formally define the competitive ratio that measures the (expected) effective rate of the suggested codes in term of the optimal (expected) effective rate of the clairvoyant code. For rateless codes, the expected effective rate is $\frac{k}{\tau_s(k)}$.
\begin{definition}
The competitive ratio for a message with $k$ bits and error probability $\epsilon$ is defined as 
\begin{align}\label{eq:def_CR_finite}
\CR(k,\epsilon)&= 
\sup_{\mathcal C_k:P_e\le\epsilon }\min_{s\in\mathcal S} \left(\frac{\frac{k}{\tau_s(k)}}{ \frac{k}{\inf\limits_{\mathcal C^*_k:P_e\le\epsilon } \tau^\ast_s(k)}}\right)=
\sup_{\mathcal C_k:P_e\le\epsilon }\min_{s\in\mathcal S}
\left(\frac{\inf\limits_{\mathcal C^*_k:P_e\le\epsilon }\tau^\ast_s(k)}{\tau_s(k)}\right).
\end{align}
\end{definition}
We note that the competitive ratio is always smaller than $1$ since the clairvoyant code $\mathcal C_k^\ast$ can always achieve the performance of the code $\mathcal C_k$. Thus, the competitive ratio serves as a percentage measure to indicate how close the performance is to the optimal one. The universality of the code with respect to the channel in use is reflected in the inner minimization over the states $\mathcal S$.
We also note that $\CR$ can be viewed both as a ratio of expected rates and a ratio of expected decoding times. Indeed, the rate and the (normalized) decoding time are reciprocal. We use both perspectives interchangeably in this work.

The competitive ratio guarantees are multiplicative. Alternatively, one may be interested in an additive performance guarantee such as delay or rate when compared to the optimal code. 
\begin{definition}
{The regret for a message with $k$ bits and a probability error $\epsilon$ is defined as
\begin{align}\label{eq:def_regret_finite}
\regret(k,\epsilon)&= 
\inf_{\mathcal C_k:P_e\le\epsilon }\max_{s\in\mathcal S} \left( \frac{k}{\inf\limits_{\mathcal C^*_k:P_e\le\epsilon } \tau^\ast_s(k)} - \frac{k}{\tau_s(k)}\right).
\end{align}}
\end{definition}
Note that, in contrast to $\CR(\epsilon,k)$ where we maximize the metric over codes, here we seek codes that minimize the rates difference.

In this paper, our focus is on the regime of vanishing probability of error $\epsilon$ as $k\to\infty$. This is formalized in the following definition.
\begin{definition}
{A family of channels $\mathcal W$ is $\alpha$-competitive if there exists a sequence $\epsilon_k\to0$ such that 
\begin{align}\label{eq:CR_inf}
    \limsup_{k\to\infty} \CR(k,\epsilon_k) \ge \alpha.
\end{align}
The supremum over all $\alpha$ that are competitive is referred to as the optimal competitive ratio, or alternatively the ``competitive capacity'', and is denoted by $\CR$.}
%\os{Do we want to name it as the "optimal competitive ratio" or the "competitive capacity", the names are not consistent with the theorems.}
\end{definition}
 Similarly, we can define the optimal regret.
 \begin{definition}
{A family of channels $\cW$ achieves a $\rho-\regret$ if there exists a sequence $\epsilon_k\to0$ such that 
\begin{align}
\label{eq:regret_inf}
    \liminf_{k\to\infty} \regret(k,\epsilon_k) \le \rho.
\end{align}
The infimum over all such $\rho$ is the \emph{optimal regret} and is denoted by $\regret$.}
\end{definition}
%\ml{check the consistency of max vs. sup}
% \ml{Our definition of $\mathbb{E}_s[\tau_k^\ast]$ is not very precise. ``Optimal'' is not well defined, as for different communication schemes there may be different stopping time values as a function of $k$. What is the error requirement in defining $\mathbb{E}_s[\tau_k^\ast]$? What is the expectation over? Perhaps define 
% $$
% T^*_s=\inf_{\substack{(E^*,\{D^*_i\},\{H^*_i\})_k \\ \mbox{s.t.},\ \lim_{k \to \infty}\Pr(M \neq  D^*_{\tau}(Y^{\tau})|S=s) = 0}}\left(\limsup_k\frac{\mathbb{E}_s[\tau_k^\ast]}{k}\right)
% $$
% We can then show that $T^*_s=\frac{1}{C_s}$.
% Now, define the $\alpha$-competitive criteria for $s \in \mathcal S$ as
% \begin{align}
% \inf_{\substack{(E,\{D_i\},\{H_i\})_k \\ \mbox{s.t.},\ \lim_{k \to \infty}\Pr(M \neq  D_{\tau}(Y^{\tau})|S=s) = 0}}\left(\limsup_k\left(\frac{kT^*_s}{\mathbb{E}_s[{\tau_k}]}\right)\right)\ge \alpha.
% \end{align}
% We can discuss via email or when we next speak.}

% \os{Alternatively, we can write?}
% \begin{align}
%     \CR&= \limsup_{k\to\infty} \sup \min_s \frac{\mathbb{E}_s[\tau_k^\ast]}{\mathbb{E}_s[{\tau_k}]},
% \end{align}
% where the supremum 

%\ml{Mike: up to here is the recent pass - May 2023}.

\section{Main results}
\label{sec:main}
In this section, we present a closed-form, single-letter formula for the optimal competitive ratio (and accordingly for regret).
To build intuition, we start with a rough description of a simple lower bound for the competitive ratio that corresponds to codes designed according to a single distribution. 

\subsection{Simple lower bound}
Let $p\in\mathcal P(\mathcal X)$ be a distribution over channel inputs.
Consider a code designed randomly according to the distribution $p$.
Let $X$ be a random variable with distribution $p$.
The expected rate for channel $W_s$ with such codes is roughly $I_s(X;Y)$. Thus, loosely speaking, we can achieve 
\begin{align}\label{eq:LB}
\CR\ge \max_p \min_{s \in {\mathcal S}} \frac{I_s(X;Y)}{C_s},
\end{align}
where $C_s = \max_{p(x)} I_s(X;Y)$ is the channel capacity when the encoder and the decoder know $s$. As it turns out, this lower bound is not optimal. 

We can think of the competitive setting as a broadcast channel with a single message (multicast), where the $s$'th decoder observes its channel output generated by $P_s(y|x)$. While this comparison to the broadcast channel is not completely accurate due to decoder state information as discussed previously, we add the discussion that follows to build intuition. Consider a special case with two decoders. If we use a code governed by a single input distribution $p$, then one of the decoders, e.g., the first, will likely be able to decode the message prior to the other decoder. In this case, to optimize the remaining time until the second decoder decodes, one can improve on code design by modifying the input distribution for the remaining code entries to that optimizing the second decoder's channel.\footnote{{Note that the encoder is not aware of the (random) time index in which the first decoder decodes the message, but our achievability shows that it is sufficient to switch the input distribution at roughly the expected decoding time.}} In the next section, we formalize these ideas with a characterization to the competitive capacity which optimizes over $|\mathcal S|$-distributions.

\subsection{The competitive capacity $\CR$ and optimal $\regret$.}
Let ${\bf p}=p_1,\dots,p_{|\mathcal S|}$ be $|\mathcal S|$ channel input distributions over $\mathcal X$, and let $X_j$ be the corresponding random variable with distribution $p_j$. 
In our definitions below, we also use an ordering  ${\bf s} = (s_1,\dots,s_{|\mathcal S|})$ of $\mathcal S$ which specifies the order in which decoding occurs. Our first definition of $\CR$ consists of an optimization over both ${\bf p}$ and ${\bf s}$. We later show that the optimal ordering ${\bf s}$ can be determined by ${\bf p}$, giving us an equivalent competitive capacity expression \eqref{eq:cr} as an optimization over ${\bf p}$ only.

In what follows, given ${\bf p}$ and ${\bf s}$, the variables $T_i^{\bpps}$ are defined inductively and correspond to the (normalized) decoding time of a receiver viewing the output of the $i$'th channel in $\mathcal S$ according to the order ${\bf s}$.
Let $T_{0}^{\bpps}=0$.
For general $i$, let $T_{i}^{\bpps}=\Delta_i\bpps+T_{i-1}^{\bpps}$ where the decoding-time increments $\Delta_i$ are defined by
\begin{align}\label{eq:def_increments}
\Delta_i\bpps \triangleq \frac{\{1 - \sum_{j=1}^{i-1} \Delta_j\bpps I_{s_i}(X_j;Y)\}_+}{I_{s_i}(X_i;Y)},
\end{align}
The expression above corresponds to a code designed using the input distributions ${\bf p}$, where distribution $p_1$ is used for the first $k\Delta_1$ coordinates, distribution $p_2$ is used for the next $k\Delta_2$ coordinates, and so on.
Using such a code, a receiver of channel $s_i$ will be able to decode once the corresponding cumulative mutual information $\sum_{j}k\Delta_j\bpps I_{s_i}(X_j;Y)$ exceeds $k$. 
Accordingly, $T_1^{\bpps}=\Delta_1$ is set such that $\Delta_1\bpps =\frac1{I_{s_1}(X_1;Y)}$;
$T_2^{\bpps}=\Delta_1+\Delta_2$ is set such that $\Delta_2$ is the residual normalized decoding time for $s_2$ (normalized by $k$ and the rate $I_2(X_2;Y)$ experienced by channel $s_2$ with input distribution $p_2$) and corresponds to the term $1 - \Delta_1\bpps I_{s_2}(X_1;Y)$; and
in general, $T_i^{\bpps}=\sum_{j=1}^i\Delta_j\bpps$ where $\Delta_i$ is the residual decoding time for a receiver of channel $s_i$ (normalized by $k$ and $I_{s_i}(X_{s_i};Y)$) and corresponds to the term $1 - \sum_{j=1}^{i-1} \Delta_j\bpps I_{s_i}(X_j;Y)$.
The following result formalizes the competitive capacity.

% The $i$th time $T_i^{\bpps}$ corresponds to the (normalized) time needed to decode a bit at the $i$th decoder. This can be computed as the sum of all decoding time increments as $T_i^{\bpps} = \sum_{j=1}^i\Delta_j\bpps$. The $i$th increment is computed as the ratio between the remaining information at time $T_{i-1}^{\bpps}$ divided by the "rate" experienced by at the channel $s_i$ with input distribution $p_i$. The following result formalizes the optimal competitive capacity. 

\begin{theorem}[Competitive Channel Capacity]\label{th:main}
The competitive capacity is given by
\begin{align}\label{eq:th_main_CR}
    \CR&= \max_{\bf p,\bf s} \min_{i \in [|\mathcal S|]} \frac{T^\ast_{s_i}}{\sum_{j=1}^i\Delta_j\bpps}
    = \max_{\bf p,\bf s} \min_{i \in [|\mathcal S|]} \frac{T^\ast_{s_i}}{T_i^{\bpps}},
\end{align}
where $T^\ast_s=\inf_{p_X}\frac{1}{I_s(X;Y)} = \frac1{C_s}$.
\end{theorem}
A trivial special case is when there is a single channel, i.e., $|\mathcal S|=1$ where $\CR=1$ and $p_1$ is the capacity-achieving distribution. In the general case, $\CR$ is defined as an optimization  over both ${\bf p}$ and ${\bf s}$. In what follows, we show that the optimization can be simplified with a \emph{greedy decoding order}.

For a fixed ${\bf p}=p_1,\dots,p_{|\mathcal S|}$, we define the greedy decoding order corresponding to ${\bf p}$ in which $s_1^{\bpp}$ is set to be the first channel for which a receiver is able to decode using a code governed by the distribution $p_1$.
Denote the first decoding time as $k\overline{\Delta}_1(\bf p)$.
Then, switching to the distribution $p_2$, state $s_2^{\bpp}$ is set to be the second channel for which a receiver can decode (using code design that starts with $p_1$ for $k\overline{\Delta}_1({\bf p})$ steps and then switches to $p_2$). 
Denote the second decoding time as $k(\overline{\Delta}_1({\bf p})+\overline{\Delta}_2({\bf p}))$.
In each phase, we switch to the next $p_i$ and set $s_i^{\bpp}$ to be the next channel to allow decoding. 
In what follows, we show that this natural order ${\bf s}^{({\bf p})}=s_1^{\bpp},\dots,s_{|\mathcal S|}^{\bpp}$ is indeed optimal.
Namely, given ${\bf p}$, the variables $T_i^{\bpp}=\sum_{j=1}^i\overline{\Delta}_1({\bf p})$ are defined inductively and correspond to the (normalized) decoding time of a receiver viewing the output of the $i$'th channel according to the greedy ordering. Let  $T_0^{\bpp}=0$.
For any $i$, define (inductively) $T_{i}^{\bpp}={\overline{\Delta}}_i(\bp)+T_{i-1}^{\bpp}$,
where the decoding-time increments, $\overline{\Delta}_i\bpp = T_i^{\bpp} - T_{i-1}^{\bpp}$, are defined~as

% $s_1^{\bpp}$ is set to be the first channel for which a receiver is able to decode using a code governed by the distribution $p_1$. Then, switching to the distribution $p_2$, $s_2^{\bpp}$ is set to be the second channel for which a receiver can decode (its decoding time is also affected by $p_1$). In each phase, we switch to the next $p_i$ and set $s_i^{\bpp}$ to be the next channel to allow decoding. In what follows, we show that this natural order ${\bf s}^{({\bf p})}=s_1^{\bpp},\dots,s_{|\mathcal S|}^{\bpp}$ is indeed optimal.

% For a fixed ${\bf p}=p_1,\dots,p_{|\mathcal S|}$, we can define the \emph{greedy decoding order} in which $s_1$ is set to be the first channel for which a corresponding receiver is able to decode using a code governed by the distribution $p_1$.
% Then, switching to the distribution $p_2$, $s_2$ is set to be the second channel for which a corresponding receiver can decode (this time is also affected by $p_1$). In each phase, we switch to the next $p_i$ and set $s_i$ to be the next channel to allow decoding. In what follows, we show that optimizing $\bs$ is redundant as this natural order ${\bf s}^{({\bf p})}=s_1^{\bpp},\dots,s_{|\mathcal S|}^{\bpp}$ is indeed optimal.

% More formally, let ${\bf p}=p_1,\dots,p_{|\mathcal S|}$ and define $T_0^{\bpp}=0$.
% For any $i$, define $T_{i}^{\bpp}=\overline{\Delta}_i(\bp)+T_{i-1}^{\bpp}$ (defined inductively),
% where the decoding-time increments, $\overline{\Delta}_i\bpp$, are defined as
\begin{align}\label{eq:delta_min}
\overline{\Delta}_i(\bp) \triangleq T_i^{\bpp} - T_{i-1}^{\bpp} = \min_{s \in \mathcal S \setminus \{s^{\bpp}_1,\dots,s^{\bpp}_{i-1}\}} \frac{1 - \sum_{j=1}^{i-1} \overline{\Delta}_j\bpp I_s(X_j;Y)}{I_s(X_i;Y)},
\end{align}
and $s_i^{\bpp}$ is defined (recursively) as the $\arg\min$ of \eqref{eq:delta_min} (ties are broken arbitrarily). 
{Intuitively, the numerator in the right-hand side of \eqref{eq:delta_min} can be thought of as the remaining information to be decoded, and the denominator corresponds to the information rate.}
Notice that, using the notation above and that presented before Theorem~\ref{th:main}, it holds that $\overline{\Delta}_i(\bp)=\Delta_i(\bp,\bs^{(\bp)})$ and $T_i^{\bpp} = T_i^{(\bp,\bs^{(\bp)})}$.

\begin{corollary}[Competitive Channel Capacity, revisited]\label{cor:ordering}
The competitive capacity is given by 
\begin{align}
\label{eq:cr}
    \CR&= \max_{\bf p} \min_{i \in [|\mathcal S|]} \frac{T^\ast_{s^{(\bp)}_i}}{\sum_{j=1}^i\Delta_j(\bp,\bs^{(\bp)})}= \max_{\bf p} \min_{i \in [|\mathcal S|]} \frac{T^\ast_{s^{(\bp)}_i}}{T_i^{(\bp)}},
\end{align}
where $T^\ast_s=\inf_{p_X}\frac{1}{I_s(X;Y)} = \frac1{C_s}$.
\end{corollary}
Corollary~\ref{cor:ordering} is proven in Appendix~\ref{sec:app:ordering}.
Similar to the $\CR$ measure in Corollary~\ref{cor:ordering}, we obtain a single-letter expression for $\regret$.

% \subsection{Regret and the general formulation}\label{subsec:main_general}
% Our second result is a single-letter expression for the regret problem.
\begin{theorem}[Optimal regret]\label{th:regret}
The optimal regret is given by
\begin{align}\label{eq:th_regret}
    \regret&= \min_{\bf p} \max_{i \in [|\mathcal S|]} \left(\frac1{T^\ast_{s_i^{\bpp}}} - \frac1{\sum_{j=1}^i \Delta_j(\bp,\bs^{(\bp)})} \right)=\min_{\bf p} \max_{i \in [|\mathcal S|]} \left(\frac1{T^\ast_{s_i^{\bpp}}} - \frac1{ T_j^{(\bp)}} \right),
\end{align}
where $T^\ast_s=\inf_{p_X}\frac{1}{I_s(X;Y)} = \frac1{C_s}$.
\end{theorem}
The achievability and converse proofs for Theorem~\ref{th:regret} closely follow the lines of proof given later for Theorem~\ref{th:main}. 
Thus, in the proof of Theorem~\ref{th:main}, in Sections \ref{sec:ach} and \ref{sec:con}, we briefly mention the modifications needed to prove Theorem~\ref{th:regret}. The optimal competitive ratio and regret expressions are different in the nature of their guarantees, but their solutions involve the same time increments. Nevertheless, the sequences of input distributions that attain the optimum in \eqref{eq:th_main_CR} and \eqref{eq:th_regret} can differ. Below, we show connections between the optimizing distributions for $\CR$ and $\regret$ in the asymptotic regime. 

\subsection{Connecting $\CR$ and $\regret$}

To connect the $\CR$ and $\regret$ metrics, we first introduce corresponding weighted variants.
In particular, for ${\bf w}=w_1,\dots,w_{|S|} \in (0,\infty)^{|S|}$, define the weighted version of $\CR$ as
\begin{align}\label{eq:def_CR_weights}
\CR(k,\epsilon,\bf{w})&= 
\sup_{\mathcal C_k:P_e\le\epsilon }\min_{s\in\mathcal S}\ \  w_s \cdot \left(\frac{ \inf\limits_{\mathcal C^*_k:P_e\le\epsilon } \tau^\ast_s(k)}{\tau_s(k)}\right).
\end{align}
For ${\bf r}=r_1,\dots,r_{|S|} \in (0,\infty)^{|S|}$, define the weighted version of $\regret$ as
\begin{align}\label{eq:def_regret_finite_w}
\regret(k,\epsilon,{\bf r})&= 
\inf_{\mathcal C_k:P_e\le\epsilon }\max_{s\in\mathcal S}\ \ r_{s} \cdot 
\left( \frac{k}{\inf\limits_{\mathcal C^*_k:P_e\le\epsilon } \tau^\ast_s(k)} - \frac{k}{\tau_s(k)}\right).
\end{align}
The asymptotic, weighted, competitive-ratio and regret are denoted by $\CR(\bf{w})$ and $\regret({\bf r})$ respectively and are defined following the lines of \eqref{eq:CR_inf} and \eqref{eq:regret_inf}. 
Similar to the proof of Theorem \ref{th:main}, one can show that 
\begin{align}\label{eq:CR_weight_sol}
\CR({\bf{w}})=\max_{\bpps} \min_{i\in [|\mathcal S|]}\ \ w_{s_i}\frac{T^*_{s_i}}{\sum_{j=1}^i \Delta_j\bpps},
\end{align}
and
\begin{align}\label{eq:regret_weight}
\regret({\bf r}) = \min_{\bpps} \max_{i\in [|\mathcal S|]}\ \ 
r_{s_i}\left(\frac1{T^*_{s_i}}-\frac1{\sum_{j=1}^i \Delta_j\bpps}\right)  
\end{align}
%\ml{We need to verify. I didn't go through the whole proof mentally in this context yet.}

% Now, we can obtain $\CR$ and $\regret$ as special cases of \eqref{eq:CR_weight_sol}. In particular, if we choose $\bf{w} = (1,\dots,1)$, we obtain the $\CR$. %To obtain the regret, we can choose $w_i = T^\ast_{s_i}$ so that the input distribution and the , we obtain the $\regret$. 

% For \eqref{eq:CR_weight_sol}, denote the set of optimizing $({\bf p}, {\bf s})$ in the optimization above by ${\tt CR}_{\bf w}$.
% Notice here, depending on ${\bf w}$, it may be that $\alpha^*_{\bf w}$ is greater than 1.

% For ${\bf r}=r_1,\dots,r_{|S|} \in (0,\infty)^{|S|}$, let the weighted regret be
% \begin{align}\label{eq:regret_weight}
% \regret({\bf r}) = \inf_{\bpps} \max_{i\in \mathcal S}
% r_{s_i}\left(\frac1{T^*_{s_i}}-\frac1{\sum_{j=1}^i \Delta_j\bpps}\right)    
% \end{align}
% We denote the set of optimizing $\bp$ in the optimization above by ${\tt Regret}_{\bf r}$.

We proceed to show that the competitive ratio and the regret problems are equivalent when considering their generalized weighted versions. Namely, we show that the ability to find the optimal %(or close to optimal) 
$\bpps$ in one implies the same for the other. 
Towards that end, denote by ${\tt Regret}_{\bf r}$ and ${\tt CR}_{\bf w}$ the arguments that achieve the optimum in \eqref{eq:CR_weight_sol} and \eqref{eq:regret_weight}, respectively.

\begin{proposition}\label{th:equivalent}
For every ${\bf w}$, there exists an ${\bf r}$ such that ${\tt Regret}_{\bf r} \subseteq {\tt CR}_{\bf w}$. Moreover, one can find ${\bf r}$ by solving (multiple instances) of \eqref{eq:regret_weight}. In the other direction, for every ${\bf r}$, there exists a ${\bf w}$ such that ${\tt CR}_{\bf w}\subseteq{\tt Regret}_{\bf r}$ and it can be found by solving (multiple instances) of \eqref{eq:CR_weight_sol}.
\end{proposition}
Proposition \ref{th:equivalent} suggests that one can solve a weighted regret problem using a solution to the weighted competitive ratio and vice versa. Proof of the proposition, i.e., the reductive analysis, is given in Appendix \ref{sec:app_equiv}.

\section{Examples}\label{sec:examples}
In this section, we study several examples to illustrate different aspects of the competitive approach of our problem formulation and its solution. 
%\subsection{Channels with two states}
Consider a family of channels $W_s(y|x)$ with $|\mathcal S| = 2$. In this case, the competitive ratio in Theorem \ref{th:main} can be written explicitly. A similar analysis (in a different context) appears in \cite{shulman2000static,Shulman:09}. 
Fix a distribution $p_1$ on $\mathcal X$, and let $s_1$ be such that $I_{s_1}(X;Y)\ge I_{s_2}(X;Y)$. 
Here $X$ is distributed according to $p_1$.
Let $p_2$ be the capacity-achieving distribution of channel $s_2$.
For $\bp=(p_1,p_2)$ and $\bs=(s_1,s_2)$, using \eqref{eq:def_increments}, the first decoding increment corresponding to $s_1$ is $\Delta_1 = \frac1{I_{s_1}(X;Y)}$.
This implies a competitive ratio for channel $s_1$ of $\frac{I_{s_1}(X;Y)}{C_{s_1}}$. For $s_2$, roughly speaking, our analysis shows that the fraction of message bits decoded until time $k\Delta_1$ is $\Delta_1 I_{s_2}(X;Y)$. Thus, for the remaining $1 - \Delta_1 I_{s_2}(X;Y)$ fraction, the corresponding time increment is set to $\Delta_2 = \frac{1-\Delta_1 I_{s_2}(X;Y)}{C_{s_2}}$. The overall normalized decoding time for channel $s_2$ is thus
\begin{align}\label{eq:2states_sumT}
\Delta_1 + \Delta_2
        &= \frac1{I_{s_1}(X;Y)} + \frac{ 1 - \frac1{I_{s_1}(X;Y)} I_{s_2}(X;Y)}{C_{s_2}}.
\end{align}
We can simplify the induced competitive ratio for $s_2$ as
$\frac{I_{s_1}(X;Y) }{ C_{s_2} + I_{s_1}(X;Y) -  I_{s_2}(X;Y)}$.
To conclude, denoting $p_1$ by $p$ below, the competitive ratio is 
\begin{align}
\label{eq:2_user}
    \CR&= \max_{p(x)} \min_{s_1(p)} \left\{ \frac{I_{s_1}(X;Y)}{C_{s_1}}, \frac{I_{s_1}(X;Y)}{ C_{s_2} + I_{s_1}(X;Y) -  I_{s_2}(X;Y)}\right\}, 
\end{align}
where $s_1(p)$ is the channel $s_1 \in \cS$ for which $I_{s_1}(X;Y)\ge I_{s_2}(X;Y)$ when $X$ has distribution $p$.
For the regret, \eqref{eq:2states_sumT} implies that in families of two channels,
\begin{align}\label{eq:regret_2states}
    \regret&= \min_{p(x)}\max_{s_1(p)} \left\{ C_{s_1} - I_{s_1}(X;Y), C_{s_2} - C_{s_2}  \frac{I_{s_1}(X;Y)}{C_{s_2} + I_{s_1}(X;Y) - I_{s_2}(X;Y)}\right\}.
\end{align}
\begin{figure}[t]
    \centering
\psfrag{0}[][][.9]{$0$}    
\psfrag{1}[][][.9]{$1$}
\psfrag{X}[][][.9]{$X$}
\psfrag{Y}[][][.9]{$Y$}  
\psfrag{z}[][][.9]{$z$}
\psfrag{c}[][][.9]{$1-z$}
\psfrag{s}[][][.9]{$s$}
\psfrag{d}[][][.9]{$1-s$}

% \psfrag{F}[][][.8]{$\overbrace{T_3(k)-\lfloor kT_3^{\bpps}\rfloor}$}
\includegraphics[scale=0.6]{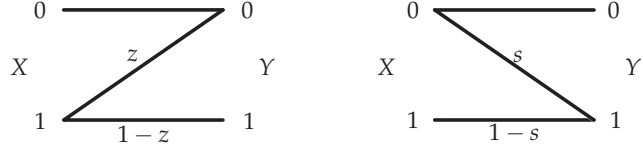}
    \caption{The Z-S channel}
    \label{fig:zs}
\end{figure}

A few concrete examples of case studies follow.

\begin{figure}[b]
    \centering
\psfrag{A}[][][.9]{(a) Lower bound }  
\psfrag{B}[][][.9]{(b) The competitive ratio}
% \psfrag{0}[][][.9]{$0$}    
% \psfrag{1}[][][.9]{$1$}
% \psfrag{X}[][][.9]{$X$}
\psfrag{Y}[][][.9]{$Y$}  
\psfrag{z}[][][.9]{$z$}
\psfrag{c}[][][.9]{$1-z$}
\psfrag{s}[][][.9]{$s$}
\includegraphics[scale=0.2]{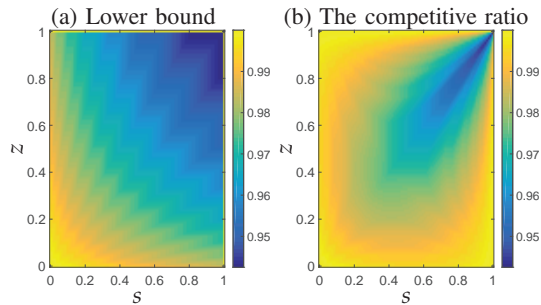}
    \caption{The competitive ratio for the Z-S channel family.  $(a)$ The lower bound induced from a single distribution. $(b)$ The optimal competitive ratio $\CR$.
    It can be noted that there is an improvement for most channel parameters.}
    \label{fig:zs_heat}
\end{figure}

\noindent
{\bf $\bullet$ The Z-S channel family:}
Consider a family of channels consisting of the Z- and the S- channels with transition parameters $z$ and $s$, respectively (see Fig. \ref{fig:zs}). 
In Fig. \ref{fig:zs_heat}, we plot in $(a)$ the lower bound on the competitive ratio based on a single (optimized) input distribution in \eqref{eq:LB}, and in (b) the optimal competitive ratio.
In both plots, the competitive ratio is in the range $[0.94,1]$. This fits the interesting result that, for binary-input channels, a uniform input distribution achieves no less than a $94\%$ of the capacity \cite{shulman2000static}. Our objective here is to illustrate the potential improvement of the optimal competitive ratio over the single input distribution lower bound in \eqref{eq:LB}. This can be observed by comparing the blue and the yellow regions in Fig. \ref{fig:zs_heat}. Next, we study channels with non-binary inputs where the  competitive ratio obtained by single-distribution codes can be significantly lower than~$\CR$. 

\noindent
{\bf $\bullet$ A channel family corresponding to a bilingual speaker:}
%\subsection{Bilingual speaker}\label{subsec:ex_bili}
The following example represents a scenario where a bilingual speaker aims to communicate with two receivers that are only able to understand one of the spoken languages \cite{shulman2000static}. 
Each receiver is modeled as a different channel. We illustrate that gains can be made due to the competitive approach with respect to the compound capacity.

Let $W_i$ be the number of symbols in each language. The channel inputs alphabet $\mathcal X=[1:W_1+W_2]$ and $|\mathcal Y_i|=[1:W_1+W_2]$ for $i=1,2$. The first channel conveys noiselessly the first $W_1$ symbols:
\begin{align}\label{eq:ex_bili}
Y_1 = \begin{cases}
X & \text{if} \ X\in\{1,\dots,W_1\}\\
Unif([W_1+1:W_1+W_2]) &\text{else}, 
\end{cases}
\end{align}
while the second channels conveys the last $W_2$ symbols  
\begin{align}
Y_2 = \begin{cases}
X & \text{if} \ X\in\{W_1+1,\dots,W_1+W_2\}\\
Unif([1:W_1]) &\text{else} 
\end{cases}
\end{align}
The capacity of the channels are $C_i = \log_2(W_i+1)$ for $i=1,2$ attained, e.g.,  with a uniform distribution over the clean symbols and one of the noisy symbols.

For $p \in [0,1]$ we consider the distribution on $\cX$ that allocates probability $p/|W_1|$ to elements of $W_1$ and $(1-p)/|W_2|$ to elements of $W_2$.
It suffices  to study such distributions as they dominate (in capacity for both the first and second channel) other distributions over $\cX$.
The achievable rates are $R_1(p) = H(p) + p \log_2(W_1)$ for the first channel and $R_2(p) = H(p) + (1-p) \log_2(W_2)$ for the second. For illustration purposes, we fix $W_1=31$ and $W_2=2$ that imply the corresponding channel capacities $C_1=5$ and $C_2 = 1.58$. 

\begin{figure}[t]
    \centering
\psfrag{E}[][][.9]{(a) Achievable rates}
\psfrag{C}[][][.9]{(b) Normalized achievable rates}
\psfrag{A}[][][.9]{(c) Normalized decoding times}  

% \psfrag{0}[][][.9]{$0$}    
% \psfrag{1}[][][.9]{$1$}
% \psfrag{X}[][][.9]{$X$}
\psfrag{F}[][][.7]{$R_i(p)$}
\psfrag{D}[][][.7]{$R_i(p)\backslash C_i$}  
\psfrag{B}[][][.7]{$T^*_i\backslash T_i(p)$}  
\psfrag{W}[r][][.7]{Compound setting}
\psfrag{L}[][][.7]{$\ \ \ \CR$ lower bound}
\psfrag{O}[][][.7]{Optimal $\CR$}
\psfrag{z}[][][.9]{}
\psfrag{c}[][][.9]{$1-z$}
\psfrag{P}[][][.9]{$p$}
\includegraphics[scale=0.5]{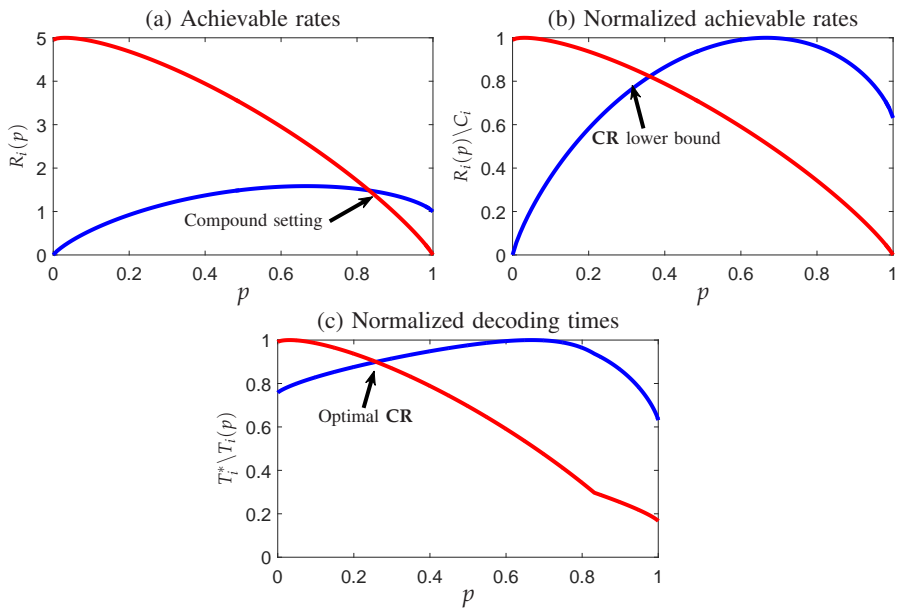}
    \caption{Illustration of the different design approaches for the Bilingual speaker example in \eqref{eq:ex_bili}. In (a), the achievable rates of the two channels are plotted as a function of the channel input parameter $p$. The max-min point corresponds to the compound channel capacity. In (b), the achievable rates normalized by their corresponding capacity show that the highest competitive ratio that can be achieved using a single input distribution rateless code is $82\%$. As a comparison, if we consider the optimal code for the compound setting with $p\sim0.83$, we obtain $94\%$ of the second (blue) channel capacity,  but only $30\%$ of the first (red) channel's capacity. In $(c)$, our optimal  competitive ratio is illustrated by looking at the normalized decoding times using two input distributions. In particular, if we use max-min point of this plot ($p\sim0.25$) until the first channel decodes and then switch to the capacity-achieving distribution of the blue channel, the competitive ratio is significantly improved to {$90\%$} and is the highest one can obtain.}
    \label{fig:HJ}
\end{figure}

In Fig. \ref{fig:HJ}, we plot in (a) the achievable rates $R_i(p)$, in (b) the achievable rates normalized by their corresponding capacities, and in (c) the competitive ratio \eqref{eq:2_user} obtained on each channel as a function of $p$. In all plots, red curves correspond to the first channel while blue curves correspond to the second channel.

In Fig. \ref{fig:HJ} (a), the max-min point corresponds to the compound channel capacity. In particular, the capacity is $R\sim 1.5$ (achieved with $p\sim0.83$).
%, and is the highest rate one can obtain if the latter is fixed with a fixed blocklength. 
In Fig. \ref{fig:HJ} (b), the achievable rates are normalized by their corresponding capacity. The max-min point here corresponds to the (single distribution)  lower bound on the competitive ratio. We note that Fig. \ref{fig:HJ} (b) implies $\CR\ge0.82$, achievable with $p=0.36$. 
That is, the max-min point guarantees that no matter what the channel is, a code generated according to $p=0.36$ achieves at least $82\%$ of the capacity for both channels. If the optimal parameter of the compound channel  $p=0.83$ was to be used, we would achieve $94\%$ out of the capacity in channel $2$ (blue curve), but this would come at the cost of achieving only $30\%$ of the first channel (red curve). 
%In other words, the compound capacity implies that first channel only used a rate of $1.5[bits/ch.]$ while the capacity is $5[bits/ch.]$.

Lastly, in Fig. \ref{fig:HJ} (c), we illustrate the optimal competitive ratio. The max-min point corresponds to the competitive ratio. {For $p \leq 0.83$, we have $R_1(p) \geq R_2(p)$ (see Fig.~\ref{fig:HJ}~(a)) and thus the first (red) channel decodes first. Thus, for $p \leq 0.83$, the red curve is the same as in Fig. \ref{fig:HJ} (b) . After the first (red) channel decodes, the capacity-achieving distribution of the second (blue) channel is used and we observe the improvement in its normalized decoding time when compared to Fig.~\ref{fig:HJ}~(b).} The opposite occurs for $p > 0.83$.
For optimal $p$, such a code achieves $90\%$ of the capacity for both channels, improving on that obtained using single-distribution codes. 
This gap can be further enlarged by increasing $W_2/W_1$. At an extreme, when both $W_1$ and $W_2/W_1$ tend to infinity, the competitive ratio corresponding to single-distribution codes approaches $0.5$ whereas that corresponding to a code designed using two distributions approaches 1.
Specifically, in this setting, $R_{1}(p)\simeq p\log{W_1} \simeq pC_1$ and $R_2(p) \simeq (1-p)\log{W_2} \simeq (1-p)C_2$.
Thus, for any single-distribution code governed by $p$, we obtain a competitive ratio of approximately $\min\{p,1-p\}$, while the optimal competitive ratio approaches 1 using the analysis in \eqref{eq:2_user}.
An alternative example (with alphabets of bounded size $4$) for which the competitive ratio of single distribution codes is 0.5  while the optimal one approaches $1$ is given next.

\noindent
{\bf $\bullet$ Modified bilingual speaker with erasures:}
In this example, we consider a modified version of the previous example. The objective is to demonstrate a setting with small input and output alphabets for which  codes designed with a single input distribution  attain a low competitive ratio while those designed with multiple (two) input distributions attain a competitive ratio that is arbitrarily close to $1$. There are two modifications from the previous example: if a channel input is used over the \emph{wrong channel},  then channel output is uniform over the entire output alphabet. The second modification is that the output of the first channel is erased with probability $1-\epsilon$. The competitive ratio is invariant to the number of symbols, and we choose $W_1=W_2=2$ for simplicity. Thus, the input alphabet is $\mathcal X= \{1,2,3,4\}$, the first channel's output alphabet is $\mathcal Y_1=\{1,2,3,4,?\}$, and that of the second channel is $\mathcal Y_2=\{1,2,3,4\}$. The first channel from $X$ to $Y_1$ is represented as a concatenation of the channel  
\begin{align}
\tilde{Y}_1 = \begin{cases}
X & \text{if} \ X\in\{1,2\}\\
Unif([1:4]) &\text{otherwise},
\end{cases}
\end{align}
and an erasure channel with erasure probability $1-\epsilon$. The output of the erasure channel is denoted by $Y_1$. The second channel in the family is 
\begin{align}
Y_2 = \begin{cases}
X & \text{if} \ X\in\{3,4\}\\
Unif([1:4]) &\text{else} 
\end{cases}
\end{align}
The capacity of the first channel is $C(s_1)=\epsilon$ and that of the second channel is $C(s_2)=1$.

By the problem symmetry, it is sufficient to compute the competitive ratio with uniform distribution over the first two inputs and the last two inputs. That is, $P(X\in\{1,2\})=\frac{p}{2}$ and $P(X \in \{ 3, 4\})=\frac{1-p}{2}$.  The lower bound on the competitive ratio using a single input distribution is
\begin{align}\label{eq:ex_HJ_e_LB}
\CR &\ge \max_p \min\{p, 1-p\}\nn\\
&= 0.5.
\end{align}
The optimal competitive ratio is computed next. For the case $p\ge \frac1{1+\epsilon}$, the channel $X\to Y_1$ decodes first and the resulting competitive ratio by \eqref{eq:2_user} is 
\begin{align}
\max_{p\ge \frac1{1+\epsilon}} \min \left\{p ,\frac{\epsilon }{1 + \epsilon } \right\}&= \frac{\epsilon }{1 + \epsilon}.
\end{align}
In the other case, $p \le \frac1{1+\epsilon}$, the second channel decodes first, and we have 
\begin{align}
\CR&=\max_{p\le \frac1{1+\epsilon}} \min \left\{1-p,\frac{1-p}{\epsilon + 1-p - \epsilon p} \right\}\\
&=\frac1{1+\epsilon}.
\end{align}
Thus, the  policy  attaining the optimal competitive ratio is $p=0$ followed by $p=1$. In other words, we utilize the capacity-achieving distributions one after the other. It is quite remarkable that this strategy achieves a ratio that is arbitrarily close to $1$ while the lower bound in \eqref{eq:ex_HJ_e_LB} can only achieve $0.5$. 

For the regret,  codes governed by a single distribution obtain
\begin{align}
    \regret&\ge \min_{p}\max\left\{\epsilon(1- p),p\right\} = \frac{\epsilon}{1+\epsilon}.
\end{align}
Considering general codes, it can be easily shown that it is optimal to decode the second channel first, which implies the constraint $p\le\frac1{1+\epsilon}$. The optimal regret is then
\begin{align}
    \regret&= \min_{p\le \frac1{1+\epsilon}}\max\left\{p,\epsilon(1 - \frac{1-p}{\epsilon + 1-p - \epsilon p})\right\} = \min_{p\le \frac1{1+\epsilon}}\max\left\{p,\frac{\epsilon^2}{1+\epsilon}\right\}=  \frac{\epsilon^2}{1+\epsilon}.
\end{align}
That is, by using two input distributions, we can improve the regret by a factor of $\epsilon$. It is also interesting to note that the optimal regret can be attained by any $p$ in the range $[0,\frac{\epsilon^2}{1+\epsilon}]$.

% \subsection{Example2: Hebrew/Japanese}
% The alphabets: $|\mathcal X|=2*W$ and $|\mathcal Y|=W$ where $W$ is the number of symbols. The "Hebrew channel" works as follows:
% \begin{align}
% Y = \begin{cases}
% X & \text{if} \ X\in\{1,\dots,W\}\\
% Unif(Y) &\text{else} 
% \end{cases}
% \end{align}
% The Japanese channel is 
% \begin{align}
% Y = \begin{cases}
% X & \text{if} \ X\in\{W+1,\dots,2W\}\\
% Unif(Y) &\text{else} 
% \end{cases}
% \end{align}
% Clearly, the capacity of each channel is $\log_2(W)$ attained with a uniform distribution over the clean symbols.

% For the lower bound, fix $p$ as the distribution allocated to the first $W$ symbols. From symmetry, we distribute $p$ among the $W$ first symbols and $1-p$ among the last $W$ symbols. For the first channel, we have $H(Y)= \log_2(W)$ and $H(Y|X) = p\log_2(W)$, and for the second $I_2(X;Y)=(1-p)\log_2(W)$. Optimizing over $p$ gives $CR=0.5$.

% For the optimal CR, again fix $p_1$as before and $p_2$ is the capacity-achieving distribution for the weaker channel (under $p_1$). We have for $p<0.5$ 
% \begin{align}
% CR &= \max_p \min\{\frac{p\log_2(W)}{\log_2(W)}, \frac{p\log_2(W)}{\log_2(W) + p \log_2(W)- (1-p)\log_2(W)}\}\\
% &=\max_p \min\{p, 0.5\}
% \end{align}
\section{Proof of Theorem~\ref{th:main}: Achievability}
\label{sec:ach}

To prove the achievability of Theorem~\ref{th:main}, below, for a given channel family, any $\alpha < \CR$, and sufficiently large $k$, we design a code ${\mathcal C}_k$ that is $\alpha$-competitive.

\begin{theorem}
\label{the:achievability}
Let $\{W_s(y|x)\}_{s \in \mathcal S}$ be a family of channels.
Let $\alpha < \CR$.
Then for any sufficiently large $k$, there exists a code ${\mathcal C}_k=\{(E,\{D_i\},\{H_i\})_k$ that is $\alpha$-competitive. 
\end{theorem}

% The proof will be presented in two steps. First we assume decoder state information (DSI). We then show how to train the decoder with only a negligible loss in the competitive ratio.

% \subsection{Proof of Theorem~\ref{the:achievability} assuming DSI}

\begin{IEEEproof}[Proof of Theorem \ref{the:achievability}]
The proof follows by applying ideas appearing, for example, in the study of finite blocklength coding theorems for discrete memoryless channels, e.g. \cite{polyanskiy2010channel}, or in the study of rateless codes in the context of channel uncertainty, e.g., \cite{tchamkerten2006variable,Shulman:09}. 
We here present proof for completeness (without trying to optimize the decoding error).
%\footnote{\ml{Perhaps add remark regarding the fact that we are only proving arbitrarily small error and add the changes needed for exponentially decaying error}}

Let ${\bf p^*}=p_1,\dots,p_{|\mathcal S|}$ and ${\bf s^{(\bp^*)}}=(s_1,\dots,s_{|\mathcal S|})$  be the optimizing distribution and ordering on ${\mathcal S}$ from the definition of $\CR$ in (\ref{eq:cr}), let $T_1,\dots,T_{|\mathcal S|}=T^{({\bf p^*})}_1,\dots,T_{|\mathcal S|}^{({\bf p^*})}$ be the corresponding normalized stopping times (we here use the notation defined above \eqref{eq:delta_min}) let $X_1,\dots,X_{|\mathcal S|}$ be the corresponding random variables, with $X_i$ distributed according to $p_i$.
%\ml{Need to prove that ${\bf p^*}$ is obtainable}.
Let $\varepsilon>0$ satisfy $\CR(1-\varepsilon)=\alpha$.
%Let $k=f_1(\varepsilon)$ and $\delta=f_2(\varepsilon,|\mathcal S|)$ for functions $f_1$ and $f_2$ to be determined shortly.
% We assume here that $T_1,\dots,T_{|\mathcal S|}$ are distinct. 
% Slight changes are needed in the analysis if this is not the case.
% Details omitted. \ml{Perhaps add details. Here, we just need to look at the distinct stopping times and map for any $s_j$ its corresponding distinct decoding time. That should formalize the analysis.}
Let $\delta>0$ be a sufficiently small constant such that $\delta < \frac{\epsilon}{4}$, and let $k$ be a sufficiently large constant to be defined shortly.

\textbf{Encoding:}
Consider a random code $\cC_k:[2^k]\rightarrow \cX^\infty$ in which for each message $m \in [2^k]$ the $i$'th coordinate $x_i(m)$ of $\cC_k(m)$ is independently distributed according to the distribution 
$\pi_i$ over $\cX$, where $\pi_i=p_j$ if $\frac{i}{k} \in [(1+\delta)T_{j-1},(1+\delta)T_{j}]$. 
Here, we define $T_0$ to be 0. The prefix of length $t$ of $\cC_k(m)$ is denoted by $x^t(m)$.
For $i > (1+\delta)T_{|\mathcal S|}k$ the codeword coordinates are set to 0. 
% Consider DSI $s \in \mathcal S$.
% Let $s=s_j$ according to the ordering above.
% For DSI $s_j \in \mathcal S$, the stopping time $\tau_i$ equals $0$ if $i<(T_j+j\delta)k$ and 1 otherwise.
% Implying that $\tau$ is deterministic and equals  $(T_j+j\delta)k=\mathbb{E}[\tau|S=s_j]$.

% \begin{lemma}
% Using the code design above, for DSI $s_j$, the typicality decoder satisfies
% \begin{align}
%     \Pr(M \neq  D_{\tau}(Y^{\tau})|S=s_j) \leq 2^{-\Omega(k)}.
% \end{align}
% Here, the probability is over code design and message $M$.
% \end{lemma}

%One may strengthen the proof to exponentially decaying error using . 
% Consider a collection of channels $\{W_s\}_{s \in \mathcal S}$.
% Consider a channel $W_s(y|x)$ and a random (infinite blocklength) codebook in which the $j$'th entry of each codeword is chosen independently according to a given distribution $p_j$ over $X$.

\textbf{Decoding:} Let $i_{i,s}(x;y)=\log\left(\frac{W_s(y|x)}{q_{i,s}(y)}\right)$ for $q_{i,s}(y)=\sum_{x \in \cX}\pi_i(x)W_s(y|x)$, and let
$i_s(x^t;y^t)=\sum_{i=1}^ti_{i,s}(x_i;y_i)$.
For every $t$, the decoder, given $y^t$, decodes a message $m$ (and stops) if there exists an $s \in \mathcal S$ such that $i_s(x^t(m);y^t) \geq k(1+\delta/2)$.
If after $t=(1+\delta)T_{|\mathcal S|}k$, no message has been decoded, the decoder stops and decodes to an arbitrary message.
We show below that such a decoder succeeds with probability $1-\frac{1}{\sqrt{k}}$ when $\sqrt{k} > \max{\left(\frac{T_{{|\mathcal{S}|}}(1+\delta)}
{T_1\delta},\frac{8T_{{|\mathcal{S}|}}(1+\delta)|\mathcal S|{\tt V}}{\delta^2}\right)}$, 
where ${\tt V}=\max_{(i,s) \in [(1+\delta)T_{|\mathcal S|}] \times {\mathcal S}}{\tt Var}[i_{i,s}(\pi_i;Y_{i,s})]$.

% It holds that 
% $E[i_s(X^t;Y^t)]=I_s^t(X;Y).$

% First, note that $\min_{s \in \mathcal S}I(P_s;Y_s)>0$, as otherwise, by our definitions, $T_s=T_{s-1}$.
% Now for $\gamma<\min_{s \in \mathcal S}I(P_s;Y_s)$ and $\delta < \epsilon/2T_s$ it holds that
% $I_s^{kT_s(1+\delta)}(X;Y) \geq k(1+2\gamma)$.
% (in our setting we may assume the collection $\{p_j\}_j$ yields a finite $T$ which is of sufficiently large value).

\textbf{Analysis:}
Let $s$ be the correct channel in use. For any $t$, let $I_s^t(X;Y)=\sum_{i=1}^tI(\pi_i;Y_{i,s})$ where the pair $(\pi_i,Y_{i,s})$ is distributed according to $\pi_i(x)W_s(y|x)$. Let $T=kT_s(1+\delta)$.
Notice that, by our definitions and setting of parameters, 
$I_s^{T}(X;Y)=(1+\delta)k$ as $T_s^{(\bp^*)}$ defined in (\ref{eq:delta_min}) is designed to satisfy $I_s^{kT_s^{(\bp^*)}}(X;Y)=k$.
A first error event, corresponding to the case in which for the transmitted message $m$ and the correct $s$,
$i_s(x^{T}(m);y^{T}) < k(1+\delta/2)$, can be bounded by above, by $\frac{1}{2|\mathcal S|\sqrt{k}}$ using, e.g., Chebyshev's inequality.
This follows from the fact that $E[i_s(X^{T};Y^{T})]=I_s^{T}(X;Y)= k(1+\delta)$ and by our assumption on~$k$.
Thus, for any chosen channel $s$ and for any message $m$
$$
\Pr_{C_k,W_s}\left[i_s(X^T(m);Y^T) < k(1+\delta/2)\right] \leq \frac{1}{2|\mathcal S|\sqrt{k}}
$$

A second error event, corresponding to the case that we decode to  $m' \ne m$, 
occurs  when for some $t \leq T$ and some $s' \in \mathcal S$, $i_{s'}(x^t(m');y^t) \geq k(1+\delta/2)$.
(Here, $s'$ is arbitrary and may be equal to the chosen channel $s$.)
Given the monotonicity of $i_{s'}(x^t(m');y^t)$ in $t$, the second error event is bounded from above by the probability that $i_{s'}(x^T(m');y^T) \geq k(1+\delta/2)$, which in turn, using the independence of $X_k^T(m')$ and $Y^T$, is bounded by
\begin{align*}
\Pr[i_{s'}(X^T(m');Y^T) \geq k(1+\delta/2)]& 
=\Pr[2^{i_{s'}(X^T(m');Y^T)} \geq 2^{k(1+\delta/2)}]\leq 2^{-k(1+\delta/2)}E\left[2^{i_{s'}(X^T(m');Y^T)}\right] \\
& = 
2^{-k(1+\delta/2)}\sum_{x^T,y^T}\left(\prod_{i=1}^T\pi_i(x_i)\right)\cdot\left(\prod_{i=1}^Tq_{i,s}(y_i)\right)\cdot \left(\prod_{i=1}^T
\frac{W_{s'}(y_i|x_i)}{q_{i,s'}(y_i)}\right) \\
& = 
2^{-k(1+\delta/2)}\sum_{y^T}\left(\prod_{i=1}^Tq_{i,s}(y_i)\right)\sum_{x^T}\left(\prod_{i=1}^T
\frac{\pi_i(x_i)W_{s'}(y_i|x_i)}{q_{i,s'}(y_i)}\right) \\
& = 
2^{-k(1+\delta/2)}\sum_{y^T}\left(\prod_{i=1}^Tq_{i,s}(y_i)\right)\prod_{i=1}^T\left(\sum_{x\in\cX}
\frac{\pi_i(x)W_{s'}(y_i|x)}{q_{i,s'}(y_i)}\right) \\
& = 
2^{-k(1+\delta/2)}\sum_{y^T}\left(\prod_{i=1}^Tq_{i,s}(y_i)\right)\cdot 1 
= 2^{-k(1+\delta/2)}.
\end{align*}
Thus, using the union bound over all $(2^k-1)|\mathcal S|$ pairs $(m',s')$, we conclude that the second error event is bounded by $|\mathcal S|2^{-k\delta/2}<\frac{1}{2|\mathcal S|\sqrt{k}}$ by our setting of parameters.
%\ml{Need to verify bound ... I added union bound on $\cS$}.
Namely,
$$
\Pr_{C_k,W_s}\left[\exists s', \exists m'\ne m, \exists t \leq T:\ \ i_{s'}(X^t(m');Y^t) \geq k(1+\delta/2)\right] \leq \frac{1}{2|\mathcal S|\sqrt{k}}.
$$
The above now implies by considering both error events and taking a union bound over $s \in \mathcal S$, that no matter which channel $s$ is in use, with probability over code design and over uniform messages $M$, successful decoding with stopping time at most ${k}T_s(1+\delta)$ holds with probability at least $1-\frac1{\sqrt{k}}$.
Namely, 
{\small{
$$
\Pr_{C_k,\{W_s\},M}\left[\forall s, \forall s', \forall m'\ne M, \forall t \leq T=kT_s(1+\delta):\ \ i_s(X^T(m);Y^T) \geq k(1+\delta/2)\ \  \text{and} \ \ i_{s'}(X^t(m');Y^t) < k(1+\delta/2)\right] \geq 1-\frac{1}{\sqrt{k}}
$$}}
This in turn implies a deterministic code with average error at most $1-\frac{1}{\sqrt{k}}$.

To conclude, we address the competitive ratio of the code at hand.
When channel $s$ is in use, the expected stopping time of the suggested scheme is at most 
\begin{align}
\label{eq:achievability}
    \tau_s(k)&= \left(1-\frac{1}{\sqrt{k}}\right)kT_s(1+\delta)+\frac1{\sqrt{k}} k T_{|\mathcal S|}(1+\delta) \leq (1+2\delta)kT_s
\end{align} (here we use the bound $\sqrt{k} > \frac{T_{|\mathcal{S}|}(1+\delta)}
{T_1\delta}$). 
% ratio apply the above analysis to our setting, notice that our encoder and equation~(\ref{eq:stopping}) 
% imply that 
% $$
% I(X,Y) = \sum_{i=1}^{j}(1+\delta)(T_i-T_{i-1})I(X_i;Y|S=s_j)k = (1+\delta)k.
% $$
% Which in turn implies that 
% $k = I(X,Y) - \frac{\delta k}{1+\delta}$;
% to obtain, with vanishing $\delta$, a vanishing decoding error.
% Namely, consider a representation of message $M \in [2^k]$ by $(M_1,\dots,M_{j})$ where for $i=1,\dots, j$, $M_i \in [2^{(T_i-T_{i-1})I(X_i;Y|S=s_j)k}]$.
% Notice that, for distinct $T_1,\dots,T_{j}$, Equation~(\ref{eq:stopping}) 
% implies that 
% $$
% k = \sum_{i=1}^{j}(T_i-T_{i-1})I(X_i;Y|S=s_j)k = \prod_{i=1}^{j} \log{|M_i|}.
% $$  
% Now, using Shannon's Channel Coding Theorem, {\em sub-message} $M_i$, that consists of $(T_i-T_{i-1})I(X_i;Y|S=s_j)k$ bits, can be communicated with success probability at least $1-2^{-\delta k}$ using entries $k[T_{i-1}+(i-1)\delta,T_{i}+i\delta]$ of the codeword corresponding to $M$.
% All in all, for sufficiently large $k$, the probability for successfully  communicating $M$ is at least $\prod_{i=1}^j(1-2^{-\delta k}) \geq 1-|\mathcal S|2^{-\delta k} \geq 1-2^{-\delta k/2}$.
%^\ml{Proof should be standard. Slackness of $\delta$ should suffice when partitioning the message to $|\mathcal S|$ sub-messages.}
As for any $s_j$ it holds that $\CR \leq \frac{T^*_{s_j}}{T_j}$,
we conclude (for our setting of $\delta$) that 
\begin{align}
\label{eq:ach_end}
    \frac{\mathbb{E}[\tau^\ast|S=s_j]}{\mathbb{E}[{\tau}|S=s_j]}=
    \frac{\tau^\ast_{s_j}(k)}{\tau_{s_j}(k)}\ge
    \frac{(1-\delta)T^*_{s_j} k}{(1+2\delta)T_j k}
    \geq \CR(1- 4\delta) > \CR(1 - \epsilon) = \alpha.
\end{align}
Here, we use the fact that for any constant $\delta>0$,  and any sequence of codes $(E^*,\{D^*_i\},\{H^*_i\})_k$ with vanishing (in $k$) decoding error, $\tau^\ast_{s_j}(k)\geq (1-\delta)T^*_{s_j} k$, which holds from Fano's inequality (see, e.g., analysis appearing later in (\ref{eq:converseT1})). 

Similar arguments to those presented above suffice to prove the achievability of Theorem~\ref{th:regret} as well.
Specifically, as in \eqref{eq:ach_end} we have for any $s_j$ and sufficiently small $\delta$ that
\begin{align*}
    \frac{k}{\tau^\ast_{s_i}(k)} - \frac{k}{\tau_{s_j}(k)}  \leq 
    \frac{1}{(1-\delta)T^*_{s_j}} -  \frac{1}{(1+2\delta)T_{j}}  \leq
    \frac{1+2\delta}{T^*_{s_j}} -  \frac{1-2\delta}{T_{j}}
    \leq
    \frac{1}{T^*_{s_j}} -  \frac{1}{T_{j}} + \frac{4\delta}{T^*_{s_j}} \leq 
    \regret + 4\delta C_{s_j}.
\end{align*}
%\ml{\rr{Does there exist a deterministic code?} I think I fixed the proof above. I will verify at a later time.}
%possible reference: %https://ieeexplore.ieee.org/stamp/stamp.jsp?tp=&arnumber=9309262
\end{IEEEproof}

\section{Proof of Theorem~\ref{th:main}: Converse}
\label{sec:con}

Let $\epsilon >0$. Let $\mathcal W=\{W_s(y|x)\}_{s \in \mathcal S}$ be a family of channels.
Consider any sequence of codes $\{\cC_k\}_k = \{(E,\{D_i\},\{H_i\})_k\}_k$ with corresponding error $\epsilon_k=P_e(\cC_k)$ that tends to zero as $k$ tends to infinity.
Let $\CR(\cC_k)$ correspond to the competitive ratio of $\cC_k$.
To prove the converse, we show that
\begin{align}
    \limsup_{k\to\infty} \CR(\cC_k) \le \CR + \epsilon.
\end{align}
More specifically, for any sufficiently large $k$ and sufficiently small corresponding $\epsilon_k$ (both may depend on $\epsilon$), we show  that $\CR(\cC_k) \le \CR + \epsilon$.

Considering $\cC_k$, recall that $\tau_s(k)= \E[\tau_k|S=s]$ is the expected decoding time for a state $s \in \cS$. Based on the decoding times, we define the decoding order of $\cC_k$ as ${\bf s}= (s_1,\dots,s_{|\mathcal S|})$ such that 
\begin{align}\label{eq:converse_ordering}
    \tau_{s_1}(k)\le \tau_{s_2}(k)\le \dots\le \tau_{s_{|\mathcal S|}}(k).
\end{align}
To simplify notation, we may use $\tau_{i}(k)$ for $\tau_{s_i}(k)$ when it is clear from the context. For any $s$, the main steps of the converse proof can be summarized as follows  
\begin{align}\label{eq:conv_main}
    \frac{\inf\limits_{\mathcal C^*_k:P_e\le\epsilon_k } \tau^\ast_s(k)}{\tau_{s}(k)}
    \stackrel{(a)}\le \frac{k T^\ast_s}{\tau_{s}(k)} + \frac{\epsilon}{2}
    \stackrel{(b)}\le \frac{ T^\ast_s}{T_s^{\bpps}} + \epsilon ,
\end{align}
where $(a)$ follows, {for sufficiently large $k$, from  \eqref{eq:achievability} in the achievability proof given in Section~\ref{sec:ach} when one considers $|S|=1$,} and $(b)$ follows for sufficiently large $k$ (through sufficiently small $\epsilon_k$) from Lemma \ref{lemma:TkleTps} below with $\bf s$ defined in \eqref{eq:converse_ordering} and $\bf p$ to be defined soon.
%Here, $\delta_k$ and $\delta'_k$ depend on $\epsilon_k$ through $\epsilon$ and will be specified later in the proof.
Thus, if the chain of inequalities above holds for all $s$, we conclude
\begin{align}
\label{eq:conv_cr}
    \CR(\cC_k) = \min_{s\in\mathcal S} \left(\frac{ \inf\limits_{\mathcal C^*_k:P_e\le\epsilon } \tau^\ast_s(k)}{\tau_s(k)}\right) &\le \min_{s\in\mathcal S} \frac{T^*_s}{T_s^{\bpps}} + \epsilon \stackrel{(a)}\le \max_{\bp,\bs}\min_{s\in\mathcal S} \frac{T^*_s}{T_s^{\bpps}} + \epsilon = \CR + \epsilon.
\end{align}
An important point to emphasize for inequality $(a)$ of \eqref{eq:conv_cr} is that $\bf p$ (defined below) does not depend on the state variable $s$ of the minimization.
% This implies that we will be able to take a maximization over which is independent of the state.}
% and $(b)$ follows from the notation $T_s(k) = \mathbb{E}_s[{\tau_k}]$ and 
% \begin{align}
% \label{eq:capacity}
% \mathbb{E}_s[{\tau_k^\ast}] \le \frac{k}{C_s}.
% \end{align}
% \ml{The above may need a formal justification - a relevant footnote should suffice}.
The regret formulation follows from a similar line of argument,
\begin{align}\label{eq:conv_main_Regret}
    \regret(\mathcal C_k)\ge \max_{s\in\cS} \left( \frac{k}{\inf\limits_{\mathcal C^*_k:P_e\le\epsilon } \tau^\ast_s(k)} - \frac{k}{\tau_s(k)}\right)\stackrel{(a)}\ge \max_{s\in\cS} \left( \frac1{T_s^\ast} - \frac{k}{\tau_s(k)}\right) - \frac{\epsilon}{2}\stackrel{(b)}\ge \max_{s\in\cS} \left( \frac1{T_s^\ast} - \frac1{T_s^{\bpps}}\right) - \epsilon,
\end{align}
where $(a)-(b)$ follow from the same argument as in \eqref{eq:conv_main}. %\os{$\delta = \frac{1}{2- T_s\epsilon}-\frac1{2}$}

The remainder of this section focuses on proving step $(b)$ in \eqref{eq:conv_main},\eqref{eq:conv_main_Regret}, which includes several challenges. We start by connecting the expected stopping time $\tau_s(k)$ induced by $\cC_k$ with $T_1^{\bpps},T_2^{\bpps},\dots$ determined by our  definition of $\bp$ described below. 

\begin{figure}[t]
    \centering
\psfrag{A}[][][.9]{$T_1(k)$}    
\psfrag{B}[][][.9]{$T_2(k)$}
\psfrag{C}[][][.9]{$T_3(k)$}
\psfrag{D}[][][.8]{{\color{red}$ \ kT_1^{\bpps}$}} 
\psfrag{E}[][][.8]{{\color{red}$ kT_2^{\bpps}$}} 
\psfrag{F}[][][.8]{{\color{red}$ kT_3^{\bpps}$}} 
\psfrag{Q}[][][.8]{$\overline{X}_1$}
\psfrag{W}[][][.8]{$\overline{X}_2$}
\psfrag{U}[][][.8]{$\overline{X}_3$}
\psfrag{R}[][][.8]{$\overline{\overline{X}}_1$}
\psfrag{T}[][][.8]{$\overline{\overline{X}}_2$}
\psfrag{Y}[][][.8]{$\overline{\overline{X}}_3$}
% \psfrag{F}[][][.8]{$\overbrace{T_3(k)-\lfloor kT_3^{\bpps}\rfloor}$}
\includegraphics[scale=0.6]{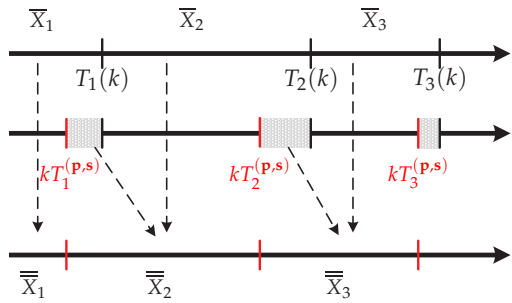}
    \caption{Illustration of Definition \ref{def:barbar}. In the first stage, the variables $\overline{X}_i$ are formed as a convex combination of certain codebook-induced distributions. %corresponding to time indices between the stopping times (that are induced by the codebook). 
    In the second stage, we form $\overline{\overline{X}}_i$ by computing the convex combinations of the grey areas with their subsequent intervals. If the grey area does not exist, then $\overline{\overline{X}}_i=\overline{X}_i$.}
    \label{fig:AR_capa}
\end{figure}

\begin{definition}\label{def:barbar}[Construction of $\bp$] 
%Let $(E,\{D_i\},\{H_i\})_k$ be a sequence of $\alpha$-competitive codes.
Consider the code $\cC_k$.
The construction f $\bp$ is done in two steps depicted in Figure~\ref{fig:AR_capa}:
\begin{enumerate}
    \item For $i \in [|\cS|]$, let $X_i$ be the distribution over $\cX$ of the $i'th$ coordinate of codewords in $\cC_k$ and let  $\overline{X}_i$ be the random variable over $\cX$ that is distributed according to the convex combination of $X_i$ from time $\tau_{i-1}(k)+1$ to $\tau_{i}(k)$. That is, 
    \begin{align}
    P(\overline{X}_i = x)&=  \frac1{\tau_{i}(k)-\tau_{i-1}(k)} \sum_{j=\tau_{i-1}(k)+1}^{\tau_{i}(k)} P(X_j=x)
    \end{align}
    \item We now recursively define random variables $\{\overline{\overline{X}}_i\}_i$ over $\cX$ with corresponding distributions $\{p_i\}$. Here, for $i=1,\dots,|\mathcal S|$, $p_i$ is the distribution of $\overline{\overline{X}}_i$.
    The first distribution $p_1$ borrows its distribution law from $\overline{X}_1$, i.e., $\overline{\overline{X}}_1 = \overline{X}_1$. The distribution $p_i$ of the $i$'th random variable $\overline{\overline{X}}_i$ is either $\overline{X}_i$ or the convex combination of $\overline{\overline{X}}_{i-1}$ and $\overline{X}_i$. Specifically, 
    \begin{align}
    \label{eq:converse:double}
    P(\overline{\overline{X}}_i = x)    &=\frac1{\tau_{i}(k)- \min\{k  T_{i-1}^{\bpps},\tau_{i-1}(k)\}} \left(\{(\tau_{i-1}(k)-  k T_{i-1}^{\bpps})\}_+\cdot  P(\overline{\overline{X}}_{i-1}=x) + (\tau_{i}(k)-\tau_{i-1}(k)) P(\overline{X}_i=x)\right)
    \end{align}
    Notice that $T_{i-1}^{\bpps}$ depends only on $p_1,\dots,p_{i-1}$ and thus \eqref{eq:converse:double} is sound.
\end{enumerate}
\end{definition}

\begin{lemma}\label{lemma:TkleTps}
For $s_i$,  
% \begin{align}\label{eq:lemma_converse_main}
% \frac{k T^\ast_s}{\mathbb{E}_s[{\tau_k}]} \le \frac{ T^\ast_s}{T_s^{\bpps}} + \frac{\epsilon}{2}
% \end{align}    
\begin{align}
\label{eq:lemma_converse_main}
     \frac{\tau_i(k)}{k}\ge T_i^{\bpps}- \gamma_i(k),
 \end{align}
where ${\bf s}=(s_1,s_2,\dots,s_{|\mathcal S|})$ is defined in \eqref{eq:converse_ordering}, $\bf p$ is defined in Definition \ref{def:barbar}, and for each $i\in[|\mathcal S|]$, $\gamma_i(k)$ tends to zero as $k$ tends to infinity.
% \ml{I need to define $\gamma_k$. At the moment $\gamma_i(k)=i\epsilon'_k$ or something like that}
\end{lemma}
\begin{IEEEproof}[Proof of Lemma \ref{lemma:TkleTps}]
We prove the lemma using an induction argument. The induction propagates through $s_1,s_2,\dots$ with two induction hypotheses -- Equation \eqref{eq:lemma_converse_main} and
\begin{align}\label{eq:lemma_converse_main_sc}
    \sum_{j=1}^{i} \left[(\tau_{j}(k)-\tau_{j-1}(k))I_{s}(\overline{X}_j;Y)\right]\le  \sum_{j=1}^{i-1} \left[ ( kT_{j}^{\bpps} - k T_{j-1}^{\bpps})I_{s}(\overline{\overline{X}}_j;Y)\right] + (\tau_{i}(k)- kT_{i-1}^{\bpps} )I_{s}(\overline{\overline{X}}_i;Y)+k \delta_i(k),
\end{align}
for all $s\in\mathcal S$.
%\rr{ and $\delta_i(k)\to0$ as $k\to\infty$}.
%, where $\delta_k\ge0$ and $\delta_k\to0$.
% \ml{Need to define $\delta_k$.}
% \ml{Need to change all $\gamma_k$ and $\delta_k$ to $\gamma_i(k)$ and $\delta_i(k)$.}\os{I think it should be clear that it goes to zero for all $i$, but I can change the notation to have this precise dependence.}
For $\epsilon_k'=\sqrt{\epsilon_k}+\frac1{k}$, let $\gamma_1(k)= T_1^{\bpps}\epsilon'_k$ and $\delta_1(k)=0$.
Let $\delta_i(k)=\gamma_{i-1}(k)\log|\mathcal Y| +\delta_{i-1}(k)$ and 
$\gamma_i(k)=\frac{\delta_{i}(k)+\epsilon'_k}{I_{s_{i}} (\overline{\overline{X}}_i;Y)}$.
As $\epsilon_k'$ tends to zero when $k$ tends to infinity, the same hold for both $\delta_i(k)$ and $\gamma_i(k)$ for every $i\in[|\mathcal S|]$.

For $s_1$, \eqref{eq:lemma_converse_main_sc} is immediately satisfied (with $\delta_1(k)=0$).
We show \eqref{eq:lemma_converse_main} using the following chain of inequalities
\begin{align}\label{eq:converseT1}
    k &=H(M)=I_{s_1}(M;Y^{\tau_{s_1}(k)})+H(M|Y^{\tau_{s_1}(k)})\nn\\
    &\stackrel{(a)}\le I_{s_1}(M;Y^{\tau_{s_1}(k)}) + k\epsilon'_k \nn\\
    &\stackrel{(b)}\le \tau_{s_1}(k) I_{s_1}(\overline{X}_1;Y) + k\epsilon'_k \nn\\
    &\stackrel{(c)}= \tau_{s_1}(k) \frac1{T_1^{\bpps}} + k\epsilon'_k,
\end{align}
where $(a)$ follows from Lemma \ref{lemma:fano} given below.
%with $\epsilon_k\to0$ as $k  \to \infty$.
% \ml{This is not precise. The value of $T_s$ in Lemma 2 is not exactly that of $T_{s_1}(k)$. We may need to modify the definition of $T_{s_1}(k)$.}\os{It is ok since in Lemma 2 we use an upper bound on $T_s$. To be more precise, in Lemma 2 the slackness that we define on $T_s$ is not an integer}
This step can be viewed as an application of Fano's inequality (modified to address expected decoding time $\tau_{s_1}(k)$).
Step $(b)$ follows from the concavity of the mutual information and Definition \ref{def:barbar}. In particular, recall by Definition \ref{def:barbar} that $\overline{X}_1$ is distributed according to uniform convex combination of the first $\tau_1(k) = \tau_{s_1}(k)$ channel inputs. Step $(c)$ follows from the definition of $T_s^{\bpps}$ in \eqref{eq:def_increments} with the first coordinate of $\bp$ taken as $p_1$ from Definition \ref{def:barbar} (and the remaining coordinates can be chosen arbitrarily). Rearranging \eqref{eq:converseT1}, we obtain \eqref{eq:lemma_converse_main} for the base case of $s_1$ with $\gamma_1(k)=T_1^{\bpps}\epsilon'_k$.

For the induction step, we assume that \eqref{eq:lemma_converse_main} and \eqref{eq:lemma_converse_main_sc} hold for $s_1,\dots,s_{i-1}$. We begin with showing \eqref{eq:lemma_converse_main_sc},
\begin{align}\label{eq:lemma_converse_main_sc_ind}
\sum_{j=1}^{i} \left[(\tau_{j}(k)-\tau_{j-1}(k))I_{s}(\overline{X}_j;Y)\right]&\stackrel{(a)}\le \sum_{j=1}^{i-2} \left[( kT_{j}^{\bpps} -  k T_{j-1}^{\bpps})I_{s}(\overline{\overline{X}}_j;Y)\right] + (\tau_{i-1}(k)- kT_{i-2}^{\bpps})I_{s}(\overline{\overline{X}}_{i-1};Y)\nn\\
&  + (\tau_{i}(k)-\tau_{i-1}(k))I_{s}(\overline{X}_{i};Y)+k\delta_{i-1}(k)\nn\\
&\stackrel{(b)}= \sum_{j=1}^{i-1} \left[( kT_{j}^{\bpps} -  k T_{j-1}^{\bpps})I_{s}(\overline{\overline{X}}_j;Y)\right] + (\tau_{i-1}(k)-  kT_{i-1}^{\bpps})I_{s}(\overline{\overline{X}}_{i-1};Y) \nn\\
& \ + (\tau_{i}(k)-\tau_{i-1}(k))I_{s}(\overline{X}_{i};Y)+k\delta_{i-1}(k)\nn\\
&\stackrel{(c)}\le \sum_{j=1}^{i-1} \left[( kT_{j}^{\bpps} -  k T_{j-1}^{\bpps})I_{s}(\overline{\overline{X}}_j;Y)\right] + (\tau_{i}(k) -  kT_{i-1}^{\bpps})I_{s}(\overline{\overline{X}}_{i};Y)\nn\\
%+ k\delta'_k
&+k\gamma_{i-1}(k)\log|\mathcal Y| +k\delta_{i-1}(k) \nn\\
&= \sum_{j=1}^{i-1} \left[( kT_{j}^{\bpps} -  k T_{j-1}^{\bpps})I_{s}(\overline{\overline{X}}_j;Y)\right] + (\tau_{i}(k) -  kT_{i-1}^{\bpps})I_{s}(\overline{\overline{X}}_{i};Y)
%+ k\delta'_k
+k\delta_i(k),
\end{align}
% \ml{Need to work out error term above}\os{$\delta_i(k)$ can be defined from the last line if we change $\gamma_k$ to $\gamma_i(k)$.}
where $(a)$ follows from the induction hypothesis. Step $(b)$ follows from splitting the term $(\tau_{i-1}(k)-  kT_{i-2}^{\bpps})$ into $(\tau_{i-1}(k)-  kT_{i-1}^{\bpps}) + (T_{i-1}^{\bpps}(k)-  kT_{i-2}^{\bpps})$. Step $(c)$ is shown based on the sign of $\tau_{i-1}(k)-  kT_{i-1}^{\bpps}$. If $\tau_{i-1}(k)-  kT_{i-1}^{\bpps}>0$, we can use the concavity of the mutual information function combined with the definition of $\overline{\overline{X}}_i$ (Definition \ref{def:barbar}) as follows
\begin{align}
(\tau_{i-1}(k)-  kT_{i-1}^{\bpps})I_{s}(\overline{\overline{X}}_{i-1};Y) + (\tau_{i}(k)-\tau_{i-1}(k))I_{s}(\overline{X}_{i};Y)&\le (\tau_{i}(k) -  kT_{i-1}^{\bpps})I_{s}(\overline{\overline{X}}_{i};Y).
\end{align}
For the other case, $\tau_{i-1}(k)-  kT_{i-1}^{\bpps}\le0$, we have
\begin{align}
    (\tau_{i-1}(k)-  kT_{i-1}^{\bpps})I_{s}(\overline{\overline{X}}_{i-1};Y) + (\tau_{i}(k)-\tau_{i-1}(k))I_{s}(\overline{X}_{i};Y)
    &= (\tau_{i-1}(k)-  kT_{i-1}^{\bpps})(I_{s}(\overline{\overline{X}}_{i-1};Y)-I_{s}(\overline{X}_{i};Y))\\
    & \ \ \ + (\tau_{i}(k)-kT_{i-1}^{\bpps})I_{s}(\overline{X}_{i};Y)\nn\\
    &= (\tau_{i-1}(k)-  kT_{i-1}^{\bpps})(I_{s}(\overline{\overline{X}}_{i-1};Y)-I_{s}(\overline{X}_{i};Y))\\
    & \ \ \ + (\tau_{i}(k)-kT_{i-1}^{\bpps})I_{s}(\overline{\overline{X}}_{i};Y)\nn\\
    &\le k\gamma_{i-1}(k)\log|\mathcal Y| + (\tau_{i}(k)-kT_{i-1}^{\bpps})I_{s}(\overline{\overline{X}}_{i};Y)
\end{align}
where the second equality follows by Definition \ref{def:barbar} that implies that $\overline{\overline{X}}_i = \overline{X}_i$ in this case, and the inequality follows from the induction hypothesis in \eqref{eq:lemma_converse_main}. 
%Step $(c)$ of \eqref{eq:lemma_converse_main_sc_ind} is completed by letting $\delta'_k = \epsilon_k\log|\mathcal Y|+\delta_k$.

To show the induction step \eqref{eq:lemma_converse_main} for $s_i$, consider
\begin{align}\label{eq:converseTi}
    k&\stackrel{(a)}\le I_{s_i}(M;Y^{\tau_{i}(k)}) + k \epsilon'_k\nn\\
    &\stackrel{(b)}\le \sum_{j=1}^{i} \left[(\tau_{j}(k)-\tau_{j-1}(k))I_{s_i}(\overline{X_j};Y) \right]+ k \epsilon'_k\nn\\ &\stackrel{(c)}\le \sum_{j=1}^{i-1} \left[( k T_{j}^{\bpps} - k T_{j-1}^{\bpps} )I_{s_i}(\overline{\overline{X}}_j;Y)\right] + (\tau_{i}(k)- kT_{i-1}^{\bpps} )I_{s_i}(\overline{\overline{X}}_i;Y) + k\delta_{i}(k)+k \epsilon'_k,
\end{align}
where $(a)-(b)$ follow from the same reasoning in \eqref{eq:converseT1}. Step $(c)$ follows from \eqref{eq:lemma_converse_main_sc_ind} above. 
Rearranging \eqref{eq:converseTi}, we obtain
\begin{align}\label{eq:converse_lemma_main_inequality}
 \frac{\tau_{i}(k)}{k}\ge T_{i-1}^{\bpps} + \frac{1 - \sum_{j=1}^{i-1} \left[(  T_{j}^{\bpps}- T_{j-1}^{\bpps} )I_{s_i}(\overline{\overline{X}}_j;Y)\right]}{I_{s_{i}} (\overline{\overline{X}}_i;Y)}- \frac{\delta_{i}(k)+\epsilon'_k}{I_{s_{i}} (\overline{\overline{X}}_i;Y)}.
\end{align}
%where $\epsilon_k''=\frac{\epsilon_k}{I_{s_{i}} (\overline{\overline{X}}_i;Y)}$.
On the other hand, recall from \eqref{eq:def_increments} that
\begin{align}\label{eq:converse_recall}
 T_i^{\bpps} &=  T_{i-1}^{\bpps} + \frac{\left\{1 -  \sum_{j=1}^{i-1} \left[(T_{j}^{\bpps}-T_{j-1}^{\bpps})I_{s_i}(X_j;Y)\right]\right\}_+ }{I_{s_i}(X_i;Y)},
\end{align}
where $\bp$ is taken from Definition \ref{def:barbar}. The proof is concluded by recalling that $\overline{\overline{X}}_i$ is distributed according to $p_i$, so we can compare directly the right hand sides of  \eqref{eq:converse_lemma_main_inequality} with  \eqref{eq:converse_recall}. 
In particular, if $\left\{1 -  \sum_{j=1}^{i-1} \left[(T_{j}^{\bpps}-T_{j-1}^{\bpps})I_{s_i}(X_j;Y)\right]\right\}_+>0$, we have 
$$
\frac{\tau_{i}(k)}{k}\ge T_{i}^{\bpps}-\frac{\delta_{i}(k)+\epsilon'_k}{I_{s_{i}} (\overline{\overline{X}}_i;Y)}=T_{i}^{\bpps}- \gamma_{i}(k).
$$
Otherwise, we have by our induction hypothesis that
\begin{align}
    \frac{\tau_{i}(k)}{k}&\ge\frac{\tau_{i-1}(k)}{k} \nn\\
    &\ge T_{i-1}^{\bpps}-\gamma_{i-1}(k)\nn\\
    &= T_{i}^{\bpps}- \gamma_{i-1}(k).
\end{align}
\end{IEEEproof}

%\ml{Up to here.}
We proceed to present the lemma that was utilized in the proof of Lemma \ref{lemma:TkleTps}. 
The lemma compares the probability of error at a random decoding time with an error probability at (roughly) the expected decoding time, and shows that the two have a negligible difference.
\begin{lemma}\label{lemma:fano}[Fano's inequality]
Consider the sequence $\{\cC_k\}$ of codes and their corresponding error bound $P_e(\cC_k) = \epsilon_k=\max_{s \in \mathcal S} \Pr(M\neq \hat{M}_\tau|S=s)$.
% , of codes and Let $(E,\{D_i\},\{H_i\})_k$ be a sequence of codes that induces an error probability $\epsilon_k(s)\triangleq \Pr(M\neq \hat{M}_\tau|S=s)$ with the random message $M$. 
For each $k$, let $\epsilon'_k = \sqrt{\epsilon_k}+ \frac1{k}$,
then $H_s(M|Y^{T_s(\epsilon_k)})\le k \epsilon'_k$ for all $s\in\mathcal S$ where $T_s(\epsilon_k) \triangleq  \tau_s(k)\frac{1+\sqrt{\epsilon_k}}{1-\sqrt{\epsilon_k}}$. 
\end{lemma}
\begin{IEEEproof}[Proof of Lemma \ref{lemma:fano}]
For a fixed $s$, we prove the claim by contradiction. With some abuse of notation, we omit the dependence of all probabilities, expectations and errors on the state $s$. Suppose $H(M|Y^{T_s(\epsilon_k)})> k\epsilon'_k$. It follows that $H(M|Y^{t})> k\epsilon'_k$ for all $t\le T_s(\epsilon_k)$. 
By the standard derivation of Fano's inequality 
\begin{align}\label{eq:fano}
    H(M|Y^t)&\le H(M,\mathbb{1}\{M\neq \hat{M}_t\}|Y^t)\nn\\
    &\le H(\mathbb{1}\{M\neq \hat{M}_t\}|Y^t) + \Pr(M\neq \hat{M}_t)\log(|\mathcal M|-1),
\end{align}
where the equality follows from the fact that the indicator $\mathbb{1}\{M\neq \hat{M}_t\}$ is a deterministic function of $(M,Y^t)$, and the inequality follows from $H(M|\mathbb{1}\{M\neq \hat{M}_t\},Y^t)\le \log (|\mathcal M|-1)$. Further, we can use \eqref{eq:fano} to lower bound the probability of error as $\Pr(M\neq \hat{M}_t) > \epsilon'_k-\frac1{k} = \sqrt{\epsilon_k}$ for all $t\le T_s(\epsilon_k)$. We can now directly show that
\begin{align}
    \epsilon_k&\ge \Pr(M\neq \widehat{M}_\tau)
    = \sum_{t=1}^\infty \Pr(\tau=t) \Pr(M\neq \widehat{M}_t)
    >  \Pr(\tau\le T_s(\epsilon_k)) \sqrt{\epsilon_k},
\end{align}
which in turn implies $\Pr(\tau>T_s(\epsilon_k)) > 1- \sqrt{\epsilon_k}$. Finally, we bound the expected decoding time as 
\begin{align}
    \tau_s(k)&\ge T_s(\epsilon_k)\Pr(\tau>T_s(\epsilon_k)) > T_s(\epsilon_k)(1- \sqrt{\epsilon_k})
    = \tau_s(k)(1+\sqrt{\epsilon_k}),
\end{align}
resulting in a contradiction.
\end{IEEEproof}

\section{Conclusions}
% We study rateless codes for communication in the face of uncertain channel statistics through the lens of competitive analysis and present single-letter expressions for the competitive-ratio and regret metrics in the asymptotic regime.
% Our model considers channels that do not vary over time and without feedback or shared randomness. The competitive analysis, in the asymptotic and/or finite-message length regime (see, \eqref{eq:def_CR_finite} and \eqref{eq:def_regret_finite}), under such extended models is an interesting subject of future studies.

We study rateless codes for communication in the face of uncertain channel statistics through the lens of competitive analysis and present single-letter expressions for the competitive-ratio and regret metrics in the asymptotic regime.
Our model considers a finite family of channels that do not vary over time and without feedback or shared randomness. The competitive analysis, in the asymptotic and/or finite-message length regime (see, \eqref{eq:def_CR_finite} and \eqref{eq:def_regret_finite}), under more general models is an interesting subject of future studies. 

\appendix

\subsection{Comparison between rateless models}
\label{sec:app_comp}
In this section we compare the achievable rate region of the ``unlimited blocklegth" rateless model used here and, e.g., in \cite{shulman2000static,Shulman:09,Blits:12} to the rate region of the ``unlimited message-length" rateless model used, e.g., in the works of \cite{eswaran2007using,eswaran2009zero,woyach2012comments,lomnitz2012communication,lomnitz2013universal}.
Loosely speaking, in the former model, communication is done over an unlimited (i.e., infinite) blocklength, while the message $M$ consists of $k$ bits.
For a channel family $\cW=\{W_s\}_{s \in \mathcal S}$, rate tuple ${\bf R}=\{R_s\}_{s \in \mathcal S}$ is achievable if there exists a coding scheme that allows successful decoding when channel $W_s$ is in use after $\frac{k}{R_s}$ channel uses.
We refer to the rate region in this case by $\cR(\cW)$.
In the latter model, again, loosely speaking, communication is done over a fixed blocklength $n$, while the message $M^\infty$ consists of a bit-string of unlimited length. 
For a channel family $\cW=\{W_s\}_{s \in \mathcal S}$, rate tuple ${\bf R^\infty}=\{R^\infty_s\}_{s \in \mathcal S}$ is achievable if when channel $W_s$ is in use, the prefix $M_1^{\ell_s}$ for $\ell_s=R_s^\infty n$ of $M^\infty$ is decoded after $n$ channel uses.
We refer to the rate region in this case by $\cR^\infty(\cW)$.
For a detailed model, see the works mentioned above.

\begin{theorem}
\label{the:comp}
$\cR^\infty(\cW) \subsetneq \cR(\cW)$.
\end{theorem}

\begin{IEEEproof}[Proof sketch of Theorem~\ref{the:comp}]
Let ${\bf R}^{\infty}=\{R^\infty_s\}_{s \in \mathcal S} \in \cR^\infty(\cW)$.
To show that ${\bf R}^{\infty} \in \cR(\cW)$, one can concatenate the $n$-blocklength code achieving  ${\bf R}^{\infty}$ in the unlimited message-length model several times in such a way that allows obtaining the rate tuple ${\bf R}=\{R_s\}_{s \in \mathcal S}$ with $R_s=R^\infty_s$ for all $s \in \mathcal S$ in the unlimited blocklength model. Below, we give an example with $|\mathcal S|=3$ that can naturally be extended to general $\mathcal S$.
Let $R_1^\infty \geq R_2^\infty \geq R_3^\infty$.
Let $k$ be a sufficiently large integer such that all expressions below are integers.
We start by using the original code $\frac{k}{R_1^\infty n}$ times.
During this process, channel $s=1$ can decode all $k$ bits, channel $s=2$ can decode a subset $k_2$ of the $k$ bits of size $k\frac{R^\infty_2}{R^\infty_1}$, and channel $s=3$ can decode a subset $k_3$ of the $k$ bits of size $k\frac{R^\infty_3}{R^\infty_1}$.
This implies that the decoding time for channel $s=1$ is $T_1 = \frac{k}{R_1^\infty}$ and that the achievable rate for $s=1$ is $R_1=R_1^\infty$ accordingly.
Note that $k_3 \subseteq k_2$.
We now focus on channels $s=2$ and $s=3$, and use the original code  $\frac{k-k_2}{R^\infty_2n}=\frac{k}{n}\left(\frac{1}{R^\infty_2}-\frac{1}{R^\infty_1}\right)$ times on the message consisting of the $k-k_2$ bits not yet decoded by channel $s=2$.
After this process, channel $s=2$ had decoded all $k$ bits, implying a total decoding time of $T_2$ of $\frac{k}{R_1^\infty}$ from the first phase and $k\left(\frac{1}{R^\infty_2}-\frac{1}{R^\infty_1}\right)$ from the second. Thus $T_2=\frac{k}{R_2^\infty}$ implying that rate $R_2=R_2^\infty$ is achieved for channel $s=2$ accordingly.
We are left with $k\left(1-\frac{R^\infty_3}{R^\infty_1}-\left(\frac{R^\infty_3 }{R^\infty_2}-\frac{R^\infty_3}{R^\infty_1}\right)\right)= k\left(1-\frac{R^\infty_3}{R^\infty_2}\right)$ bits to decode at channel $s=3$. This can be done using the original code $\frac{k}{R_3^\infty n}\left(1-\frac{R^\infty_3}{R^\infty_2}\right)$ times applied to the bits not yet decoded at channel $s=3$.
All in all, channel $s=3$ decodes after $T_3=\frac{k}{R_1^\infty}+K\left(\frac{1}{R^\infty_2}-\frac{1}{R^\infty_1}\right)+k\left(\frac{1}{R^\infty_3}-\frac{1}{R^\infty_2}\right)=\frac{k}{R^\infty_3}$ implying a corresponding achievable rate of $R_3=R^\infty_3$ for channel $s=3$. The general case with $|\mathcal S|$ channels can be derived similarly.

To show that the inclusion is tight, namely, that not all ${\bf R} \in \cR(\cW) $ are in $\cR^\infty(\cW)$, consider the bilingual speaker example from Section~\ref{sec:examples}. Denote the channel family by $\cW$.
It can be shown that the rate tuple ${\bf R}=\left(5,1.2\right)$ is in $\cR(\cW)$. However, roughly speaking, to obtain ${\bf R^\infty} = (R_1^\infty,R_2^\infty) \in \cR^\infty(\cW)$ with $R_1^\infty=5$, one must use codebooks in which the elements $\{W_1+1,\dots,W_1+W_2\}$ appear {\em sparsely}, implying that $R_2^\infty \leq 1$.
Thus, ${\bf R^\infty}=\left(5,1.2\right) \not \in \cR^\infty(\cW)$.
\end{IEEEproof}

\subsection{Proof of Corollary \ref{cor:ordering}}
\label{sec:app:ordering}

\begin{IEEEproof}
For any distribution sequence $\bp$ and order $\bs$, we present the existence of another distribution $\bp^*$ such that $\CR(\bp^*,\bs^{(\bp^*)}) \geq \CR(\bp,\bs)$.
Here,  $\CR(\bp,\bs) = \min_i \frac{T^\ast_{s_i}}{\sum_{j=1}^i\Delta_j\bpps}$ and by \eqref{eq:th_main_CR}, $\CR = \max_{\bp,\bs}\CR(\bp,\bs)$.
This suffices to prove the corollary.

Let $\cS=\{1,2,\dots,|{\mathcal S}|\}$.
Consider any pair $\bp=p_1,\dots,p_{|\mathcal S|}$ and 
$\bs=s_1,\dots,s_{|\mathcal S|}$.
Assume that $\bs^{(\bp)} = (1,2,\dots,|\mathcal S|)$ is the order induced by the greedy choice of decoding.
Let $i$ be the minimal index for which $s_i \ne i$. Let $k>i$ be the location of $i$ in $\bs$, i.e., $s_k=i$.
Notice, that this implies $\Delta_k(\bp,\bs)=0$, as for channel $i$ we have 
$\Delta_i(\bp,\bs^{(\bp)}) \leq \Delta_i(\bp,\bs)$ due to the greedy nature of $\bs^{(\bp)}$.
Define a new ordering $\bs'=s'_1,\dots,s'_{\mathcal S}$ that is obtained by setting $s'_j=s_j$ for $j < i$, $s'_i=i$,  $s'_{j+1}=s_{j}$ for $i \leq j < k$, and $s'_{j}=s_{j}$ for $j > k$. 
Namely, $\bs'$ is obtained from $\bs$ by inserting $s'_i=i$ and shifting the remaining entries of $\bs'$ to the right.
For example, $\bs = (1,3,5,2,4)$ translates to $\bs' = (1,2,3,5,4)$. 
Moreover, define a corresponding $\bp'$ in which $p'_j=p_j$ for $j \leq i$, 
$p'_{i+1}=p_i$, 
$p'_j=p_{j-1}$ for $i+2 \le j\le k$, and $p'_j=p_{j}$ for $j>k$.
Namely, $\bp'$ is obtained from $\bp$ by duplicating $p_i$ in locations $p'_i=p'_{i+1}=p_i$ and by removing $p_k$ from $\bp'$. 
For our example, if $\bp=p_1,p_2,p_3,p_4,p_5$ then $\bp'=p_1,p_2,p_2,p_3,p_5$.

It now holds by our definitions that 
$\CR(\bp',\bs') \geq \CR(\bp,\bs)$.
Specifically, for each channel $s \ne i$, let $\ell(s)$ be its location in $\bs$ (i.e., $s_{\ell(s)}=s$) and $\ell'(s)$ be its location in $\bs'$ (i.e., $s'_{\ell'(s)}=s$). Then,
$
\sum_{j=1}^{\ell(s)}\Delta_j(\bp,\bs) = 
\sum_{j=1}^{\ell'(s)}\Delta_j(\bp',\bs').
$
Moreover, for the channel $s=i$,
$
\sum_{j=1}^{k}\Delta_j(\bp,\bs) \geq 
\sum_{j=1}^{i}\Delta_j(\bp',\bs').
$

We conclude that, after starting from the pair $(\bp,\bs)$ in which $i$ was the first entry for which $s_i \ne s^{(\bp)}_i$, we designed a new pair $(\bp',\bs')$ for which the first entry $i'$ (if such exists) in which $s'_{i'} \ne s^{(\bp')}_{i'}$ satisfies $i'>i$. Moreover, $\CR(\bp',\bs') \geq \CR(\bp,\bs)$.
We can now continue with $(\bp',\bs')$ and perform the process above again, or in general multiple times, each time increasing the value of $i$ for which $s_i \ne s^{(\bp)}_i$ (for the current $\bs$), and increasing the value of the corresponding competitive ratio, until we obtain a pair $(\bp^*,\bs^*)$ for which $\bs^*=\bs^{(\bp^*)}$ and 
$\CR(\bp^*,\bs^{(\bp^*)}) \geq \CR(\bp,\bs)$ as asserted.
This concludes our proof.
\end{IEEEproof}

\subsection{Proof of Proposition \ref{th:equivalent}}\label{sec:app_equiv}

\section{Relating $\CR(\bf w)$ and $\regret(\bf{r})$ (Proof of Proposition \ref{th:equivalent})}
\begin{IEEEproof}
Assume that $\regret>0$, as otherwise the assertion is immediate.
Let ${\bf w}=w_1,\dots,w_{|S|} \in (0,\infty)^{|S|}$.
% We can assume w.l.o.g. that $w_i \leq 1$ as ${\tt CR_{{\bf w}}} = {\tt CR}_{\theta{\bf w}}$ for any $\theta>0$.
%Let ${\bf r}$ be defined by
%$r^\alpha_j=\frac{\alpha T^*_j}{w_j-\alpha}$.
% $r_j=w_j T^*_j$.
% It now holds that the term in 
% $\regret({\bf r})$ is
% \begin{align*}
% r_{s_j}\left(\frac1{T^*_{s_j}}-\frac1{T^{\bpps}_{s_j}}\right) = 
% w_{s_j}\left(1-\frac{T^*_{s_j}}{T^{\bpps}_{s_j}}\right).
% \end{align*}
% Thus, any $({\bf p}, {\bf s}) \in {\tt Regret_{{\bf r}}}$.
% which can be re-arranged as
% \begin{align*}
% w_{s_j}\frac{T^*_{s_j}}{T^{({\bf p}, {\bf s})}_{s_j}} \geq \alpha^*.
% \end{align*}
For $\alpha \geq 0$ let ${\bf r}^\alpha$ be defined by
%$r^\alpha_j=\frac{\alpha T^*_j}{w_j-\alpha}$.
$r^\alpha_j=\frac{w_j T^*_j}{w_j-\alpha}$.
%\cdot \frac{1}{\max\{T^*_1,\dots,T^*_{|\mathcal S|}\}}$.
Denote the value of $\regret({\bf r}^\alpha)$ by $\rho^*_{{\bf r}^\alpha}$.
As ${\bf r}^\alpha$ is monotone increasing with $\alpha$ (in each coordinate) it follows that $\rho^*_{{\bf r}^\alpha}$ is increasing with $\alpha$ as well.
Using the continuity of $\rho^*_{{\bf r}^\alpha}$ as a function of $\alpha$, let $\alpha^*$ satisfy $\rho^*_{{\bf r}^{\alpha^*}}=1$, as $\alpha=0$ implies $\rho^*_{{\bf r}^\alpha} \leq 1$ and $\alpha\to \min_j w_j$ implies $\rho^*_{{\bf r}^\alpha}\to\infty$.
To simplify notation, denote ${\bf r}^{\alpha^*}$ by ${\bf r}$.
It now follows for any $({\bf p}, {\bf s}) \in {\tt Regret_{{\bf r}}}$ and any $j \in [|S|]$ that
\begin{align*}
r_{s_j}\left(\frac1{T^*_{s_j}}-\frac1{T^{\bpps}_{s_j}}\right) = 
\frac{w_{s_j}T^*_{s_j}}{w_{s_j}-\alpha^\ast}\left(\frac1{T^*_{s_j}}-\frac1{T^{\bpps}_{s_j}}\right) \leq \rho^*_{\bf r} = 1
\end{align*}
which can be re-arranged as
\begin{align*}
w_{s_j}\frac{T^*_{s_j}}{T^{({\bf p}, {\bf s})}_{s_j}} \geq \alpha^*.
\end{align*}
To conclude the first part of our assertion, we need to show that $\alpha^*=\CR({\bf w})$ and thus $({\bf p}, {\bf s}) \in {\tt CR_{{\bf w}}}$.
Assume by contradiction that $\alpha^*$ is strictly less than $\CR({\bf w})$.
Then there exists $({\bf p_w}, {\bf s_w}) \in {\tt CR_{{\bf w}}}$ such that 
\begin{align*}
\min_j w_{{(s_w)}_j}\frac{T^*_{{(s_w)}_j}}{T^{({\bf p_w}, {\bf s_w})}_{{(s_w)}_j}} \geq \CR({\bf w}) > \alpha^*.
\end{align*}
However, using the equivalences above, this implies for all $j$ that 
\begin{align*}
r_{{(s_w)}_j}\left(\frac1{T^*_{{(s_w)}_j}}-\frac1{T^{({\bf p_w},
{\bf s_w})}_{{(s_w)}_j}}\right) < 1
\end{align*}
In contradiction to the fact that $\rho^*_{\bf r} = 1$.
This concludes the determination of $\CR({\bf w})$ and optimizing $({\bf p}, {\bf s}) \in {\tt CR_{{\bf w}}}$ using the solution for $\regret({\bf r})$ (multiple times, until finding the correct $\alpha$ for which $\rho^*_{{\bf r}^\alpha}=1$).

The other direction of the proof works similarly by choosing  weights  $w_j^\rho =\frac{r_j}{r_j-\rho T^*_j}$ for $\rho>0$.
\end{IEEEproof}

\bibliographystyle{unsrt}
\bibliography{arxiv1}

\end{document}